\documentclass[journal=jctcce,manuscript=article]{achemso}

\usepackage{chemformula}
\usepackage{graphicx}
\usepackage{dcolumn}
\usepackage{bm}
\usepackage{lipsum}
\usepackage[utf8]{inputenc}
\usepackage[T1]{fontenc}
\usepackage{mathptmx}
\usepackage{dblfloatfix}
\usepackage{amsmath}
\usepackage{amssymb}

\mciteErrorOnUnknownfalse

\title{General correlated geminal ansatz for electronic structure calculations:\\ exploiting Pfaffians in place of determinants}

\author{Claudio Genovese}
\email{claudio.genovese@sissa.it}
\affiliation[SISSA]{SISSA, International School for Advanced Studies, Via Bonomea 265, 34136 Trieste, Italy;}

\author{Tomonori Shirakawa}
\affiliation[RIKEN]{Computational Materials Science Research Team, RIKEN Center for Computational Science (R-CCS), Hyogo 650-0047, Japan}

\author{Kousuke Nakano}
\affiliation[SISSA]{SISSA, International School for Advanced Studies, Via Bonomea 265, 34136 Trieste, Italy;}
 \alsoaffiliation[JAIST]{School of Information Science, Japan Advanced Institute of Science and Technology (JAIST), Asahidai 1-1, Nomi, Ishikawa 923-1292, Japan;}

\author{Sandro Sorella}
\email{sorella@sissa.it}
\affiliation[SISSA]{SISSA, International School for Advanced Studies, Via Bonomea 265, 34136 Trieste, Italy;}

\date{\today}
\begin{document}

\begin{abstract}
We propose here a single Pfaffian correlated variational ansatz,
 that dramatically improves the accuracy with respect to the single determinant 
one, while remaining at a similar computational cost.  
 A much larger  correlation energy is indeed determined by the most general two electron pairing function, including both singlet and triplet channels, combined with a many-body Jastrow  factor, including all possible spin-spin spin-density and density-density terms.  The main technical ingredient to exploit this accuracy  
 is the use of the Pfaffian for antisymmetrizing an highly correlated 
pairing function, thus recovering the Fermi statistics for electrons 
with an affordable computational cost. 
Moreover the application of the Diffusion Monte Carlo, within the 
fixed node approximation, allows us to obtain very accurate binding energies for the first 
preliminary calculations reported in this study: C$_2$, N$_2$ and O$_2$ and the benzene molecule. This is promising and remarkable, considering that they represent extremely difficult molecules  even for computationally demanding multi-determinant approaches, and opens therefore the way for  realistic and accurate electronic  simulations with an algorithm scaling at most as the fourth power of the number of electrons.
\end{abstract}

\maketitle

\begin{figure}[h!]
        \centering
        \includegraphics[scale=0.25]{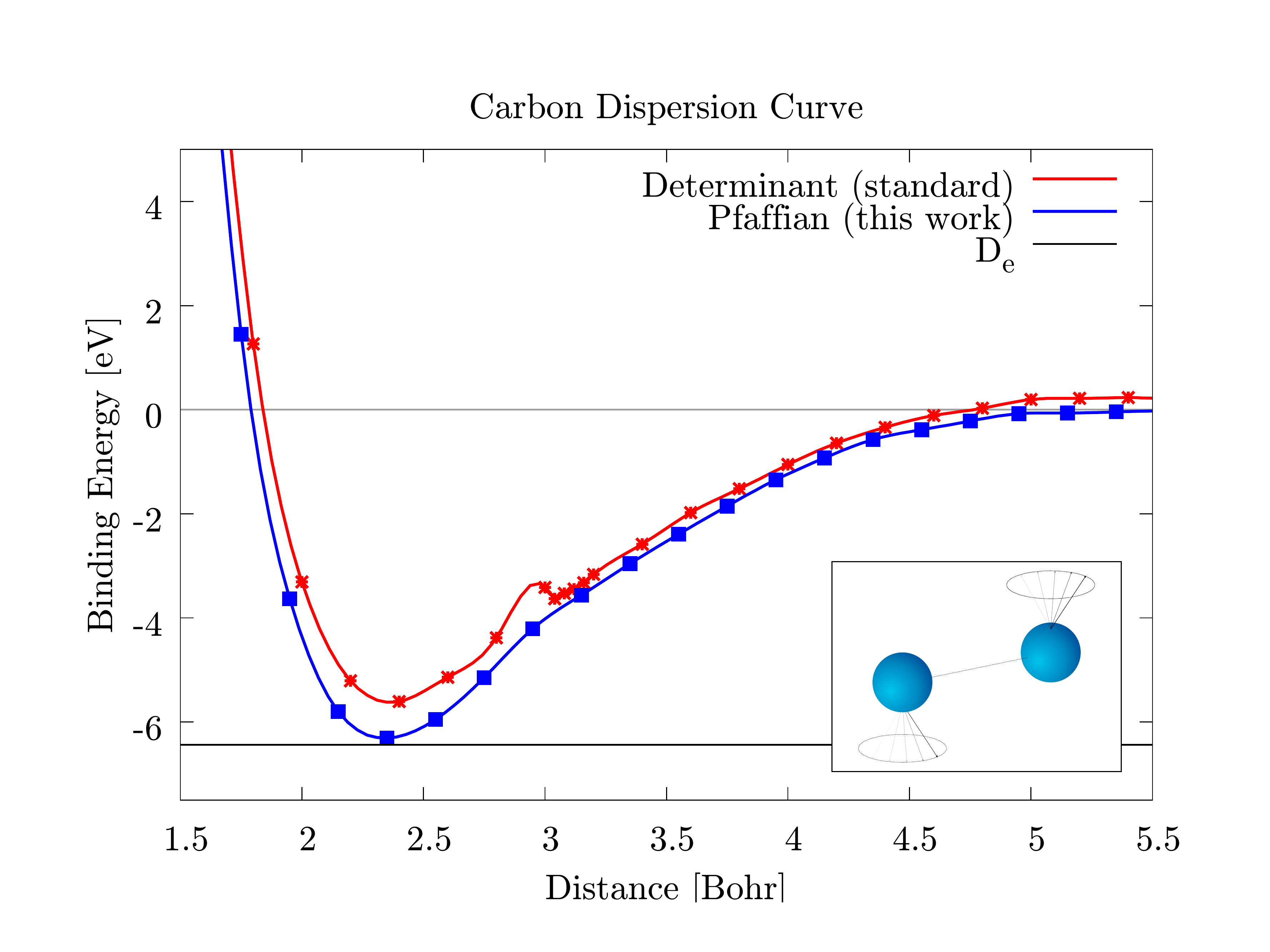}
\end{figure}

\section{Introduction}

The accurate determination  of the many-electron wave function has always been a challenging task starting from the early stage of quantum mechanics \cite{10.2307/95222}. So far several attempts are moving toward this direction ranging from CCSD(T)\cite{doi:10.1063/1.1727484} to  tensor network\cite{orus2019} and DMRG\cite{PhysRevLett.69.2863}, up to the very recent breakthrough with the use 
of machine learning methodologies\cite{pfau2019abinitio}. All these schemes, pay the price of being computationally demanding, with a computational complexity ranging from a large degree polynomial of  the number 
of electrons,  to  exponential complexity.

Quantum Monte Carlo (QMC) techniques for electronic structure calculations have proven to be very  successful in describing the electronic 
correlation encoded in a many-body wave function (WF)\cite{Booth2013,  PhysRevX.7.031059, motta2019groundstate}. In particular the variational Monte Carlo (VMC)\cite{RevModPhys, becca_sorella_2017, citalibro} samples the real space electronic configurations of the considered system with a probability distribution given by the WF square,  thus providing the efficient evaluation not only of  the total energy, but also of the  expectation values of most commonly used  many-body operators. Within VMC, it is possible to improve the description of the ground state (GS) WF by minimization of the total energy expectation value. The WF obtained can be used as it is or further improved by the diffusion Monte Carlo calculation (DMC) method \cite{doi:10.1063/1.4822960, citalibro, becca_sorella_2017, doi:10.1063/1.431514}. This technique is a projection algorithm performed statistically using the information on the sign contained in the given WF, dubbed here as guiding function. In this way we can considerably improve the description of the GS, projecting on the lowest possible energy WF with the same signs of the guiding WF. In the ideal case of a guiding function that, for  every configuration, has the same sign of the GS, the 
above described DMC algorithm provides  the exact solution\cite{citalibro, becca_sorella_2017}.

In the framework of QMC different {\it ansatzs} are used to approximate the true GS WF, with the purpose to achieve  an affordable compromise  between the accuracy of the calculation and its computational cost. Though  a good representation of the GS can sometimes be achieved with a simple and "cheap" WF, in most cases the  use of a very complicated and computationally demanding {\it ansatz} is necessary to achieve a correct  answer.

Slater determinants (SD) are the simplest fermionic WF used for QMC. They provide a single particle picture of the quantum many-body problem,   preserving the Pauli principle, i.e. the Fermi statistics for 
electrons. They can be obtained directly from mean field calculations. Unfortunately in many situations of interest it is not possible to give a good description of the system in terms of a SD\cite{doi:10.1063/1.5081933, doi:10.1063/1.4986216, doi:10.1063/1.3555821, doi:10.1063/1.2908237}. In QMC there are two possible strategies to  overcome this problem: the use of a linear combination of different SDs \cite{doi:10.1063/1.2908237,KNOWLES1984315, doi:10.1063/1.455063, ROOS1980157, doi:10.1063/1.471865} or different {\it ansatzs} with larger variational freedom\cite{doi:10.1002/ijch.198000009,  PhysRevLett.96.130201}. The multi determinant WFs can be systematically improved and in principle can describe exactly every GS with a large enough number of SDs. Unfortunately the number of SDs that have to be taken into account scales exponentially with the number of electrons preventing the calculations on large systems\cite{doi:10.1021/ct3003404}. 

The use of pairing function replaces the single particle description of the SD with a richer one in terms of electron pairs. The corresponding  WF is a natural extension of the SD {\it ansatz } and represents a direct and efficient implementation of the Anderson resonating valence bond  (RVB)\cite{Anderson:1975} theory  of many-electron WFs. In particular it provides a direct description of the singlet and triplet correlations that are absent in the SD. Depending on the definition of the pairing function, qualitatively  different WFs can be obtained. They will be described  in Sec.II, where we will focus also on the technical  details required for the calculation. We will introduce the symmetric Antisymmetrized Geminal Power (AGPs)\cite{doi:10.1002/ijch.198000009, doi:10.1063/1.1604379, doi:10.1063/1.1794632} and the broken symmetry Antisymmetrized Geminal Power (AGPu)\cite{doi:10.1021/acs.jctc.6b00288},  but we will mainly focus on the most general Antisymmetrized Geminal Power (AGP). In the previous literature \cite{PhysRevLett.96.130201,PhysRevB.77.115112} it has been indicated as Pfaffian WF, and people has been referring to the AGPs as AGP, but since the AGP (or Pfaffian WF) literally realizes the most general Antisymmetrized Geminal Power we dub this case  
with the shortest acronym, i.e. AGP. It will be  shown that this approach becomes very efficient in combination with an explicit correlation term, known as Jastrow factor (JF)\cite{PhysRevLett.109.203001, doi:10.1063/1.4829835, doi:10.1063/1.5081933}, that promotes or penalizes the bonds according to the electronic correlation. As it will be shown later,   
we have introduced a quite  general  JF, depending both on spin and electron charges. When it is applied  to an AGP without definite spin, it allows  its almost complete restoration,  mimicking in this way 
a spin projection operation that, though approximate,  is much cheaper than other approaches \cite{doi:10.1021/acs.jpca.9b01583, PhysRevA.96.022506}.  Even if the pairing functions can not be improved systematically, these WFs have a much larger variational freedom than the SD with a similar computational cost. 

If on one hand,  for the multi determinant WF, the calculation can be computationally very expensive, on the other hand, for the pairing functions,  the optimization 
of a large number of non linear 
variational parameters can become a serious limitation if not handled efficiently. Indeed, in order to exploit the full potential of these {\it ansatz} it has been fundamental to use the most recent techniques for the calculation of the derivatives and optimization strategies.

In a previous attempt, the  AGP WF was used by exploiting only a very small fraction of the large variational freedom of the {\it ansatz }\cite{PhysRevLett.96.130201,PhysRevB.77.115112}. The results were not encouraging and the energies obtained with this {\it ansatz }did not improve the ones of the AGPs that, in turns, has a lower computational cost. 
Despite the Pfaffian was no longer used in electronic system to our knowledge,
 the experience with lattice models have  shown that the  AGP WF is able to improve considerably the description of magnetic and correlated systems\cite{10.1143/PTP.97.399}.
 Moreover, the introduction of a powerful JF and the recent results obtained in combination with the AGPs \cite{PhysRevLett.121.066402, doi:10.1063/1.5081933, doi:10.1021/acs.jctc.9b00295} encouraged us to look for the unexpressed potential of the full AGP WF.

In this paper we will compare the results obtained with AGP WF with available state of the art VMC and DMC calculations. In particular we benchmark  our WF
on the diatomic molecules with corresponding high spin atoms in the first row of the periodic table, carbon, nitrogen and oxygen, and on the benzene. The first ones are systems that, despite their apparent simplicity, represent useful benchmarks for many highly correlated methods\cite{doi:10.1063/1.462649,cc2008,cc2016, doi:10.1063/1.3288054}. We will show that with the use of our best WFs, even with a very compact basis set, we are able to achieve an accuracy comparable with the state of the art multi determinant WFs at a computational cost similar to the one of a single SD. Not only the total energies and the dissociation energies are extremely accurate, but we also analyzed the magnetic proprieties of these molecules unveiling part of the rich physics behind these systems. Finally we consider  the benzene molecule, a system that represents the prototypical example of the RVB theory and thus a fundamental test case for our approach.

\section{Wave functions and procedure}
\label{WFmethod}

For all the calculations we present in this paper we used the TurboRVB package for QMC calculations \cite{ourreview, webpage}.  The WFs used for this work are factorized as the product of a fermionic mean field and an explicit bosonic correlation factor. Being $\Psi(\mathbf{X})$ the WF of a given configuration $\mathbf{X}=(\mathbf {r}_1 \sigma_1,\mathbf {r}_2 \sigma_2,\dots, \mathbf {r}_N \sigma_N)$ of $N$ electrons of spins $\sigma_i$ and positions $\mathbf{r}_i$, we can write $\Psi(\mathbf{X})$ as 
\begin{equation}
  \Psi(\mathbf{X})=J(\mathbf{X}) \times \Phi(\mathbf{X}),
  \label{jdet}
\end{equation}
where $\Phi(\mathbf{X})$ takes into account the fermionic nature of the electrons, while $J(\mathbf{X})$ 
is the JF: an exponential modulation of the WF that  substantially improves the electronic correlation description for all types of WFs
 studied here. 
The fermionic term of the WF, dubbed as $\Phi(\mathbf{X})$ in Eq.~(\ref{jdet}), is the most important part, directly encoding the behaviour of the electrons while imposing the antisymmetrization under particles exchange. In the following we will describe the basis set used, the definition of the AGPs, AGPu and AGP after a brief introduction to the Slater determinant (SD).  Finally we will discuss the JF correlator.

\subsection{Basis Set}\label{sec:basis}

 We expand our ansatz  in an atom-centered basis set of gaussian orbitals for the calculation of the JF and a hybrid basis set for the fermionic part of the WF, as it  will be discussed   in the following. 
The gaussian orbitals basis set is indicated as $\left \{ \phi_{I, \nu}(\mathbf{r})  \right \}$, with each element being the $\nu$-th orbital centred on the $I$-th atom at the position $\mathbf{R}_I$. The elements in the basis set have the  form
\begin{equation}
  \phi_{I,\nu}(\mathbf{r})\propto e^{- Z_\nu |\mathbf{r}-\mathbf{R_I}|^2} [Y_{l_\nu,m_\nu}\pm Y_{l_\nu,-m_\nu}],
  \label{basis}
\end{equation}
where $Z_\nu$ is a numerical coefficient that describes how diffuse the atomic orbital is around the atom, while $Y_{l_\nu,m_\mu}$ is the spherical harmonic function with angular quantum numbers $l_\nu$ and $m_\nu$ corresponding  to the orbital type  $\nu$ which is always assumed to be real. This basis set has been used without further contractions for the description of the JF. Instead, for the fermionic part of our WF we have  used hybrid atomic orbitals (HO)\cite{doi:10.1063/1.1794632, doi:10.1063/1.1604379} to expand them   over a richer set of gaussian orbitals and,  by means of 
the contraction, remaining  with an  affordable  number of variational parameters. The HOs, indeed, are obtained as linear combinations of all the elements of the gaussian basis set used for 
a given atom, labeled by $I$:
\begin{equation}
  	\bar{\phi}_{I,\omega}(\mathbf{r})=\sum_\nu \mu_{\omega, \nu} \phi_{I,\nu}(\mathbf{r}).
	 \label{hyb-basis}
\end{equation}

The  above  hybrid orbitals allow us  to take into account the modification 
of the standard Slater orbitals corresponding to isolated atoms, to the case when they 
are instead placed in a complex environment. Therefore we  have  chosen
to use a  number  of hybrid orbitals equal to  the 
 single  particle ones occupied in absence of electron-electron interaction 
 and including also all the ones corresponding to same shell of degenerate one particle levels.
The  corresponding orbitals 
are the ones that should physically play a role in the considered electronic systems. Hence in all the first row molecules we have considered the full hybridization of  five  atomic orbitals, coming from  two $s-$wave  and three $p-$wave ones, 
that can be corrected by several components with much higher angular 
momenta. This  is because  the full spherical  symmetry is no longer satisfied even in a simple homonuclear molecule. For the sake of compactness we indicate in the following all the basis elements  as $\left \{\phi_{k}(\mathbf{r})  \right \}$ combining the indices $\omega$ and $I$, and $I$ and $\nu$ in a single index $k$ for a lighter notation. Every time we refer to the AGPs, AGPu and AGP the basis is meant to be a basis of HOs. 

The exponents $Z_\nu$ have been chosen from the ccpVDZ or ccpVTZ basis set according to this criterium: the contraction are removed and all the exponents with  $Z_\nu>150\ {a.u}^{-1}$ are eliminated. This is possible because contracted orbitals containing very large exponents are necessary only with a  pure Gaussian basis in order  to satisfy the electron-ion cusp conditions, that is instead appropriately considered  by the one-body term of our JF, as described in sec.\ref{sec:jf}. The exponents chosen are then further optimized   at molecular equilibrium  distance and kept  fixed in the corresponding  
atomic calculation (where the optimization of the  exponents has an almost 
negligible effect) and the dispersion energy curves.

\subsection{The Slater Determinant} 

In the following we will provide a preparation description of the SD, that is important 
both for the initialization of the pairing function, and for comparing our results with existing  literature\cite{doi:10.1063/1.2908237}. From a theoretical and computational point of view the simplest fermionic WF is the Slater determinant (SD), called Jastrow SD (JSD) in the presence of a JF. The SD is built from the vacuum state by populating a number of orthogonal single particle molecular orbitals (MO) equal to the number of electrons in the system. 
Henceforth,  we omit the spin indices, by assuming that to each spin component corresponds a different Slater determinant. In our basis the MO are in the form
\begin{equation}
  \Phi^{mol}_\alpha(\mathbf{r})=\sum_{k=1}P_{\alpha,k}\phi_k(\mathbf{r}).
  \label{molorb}
\end{equation}
The MOs can be obtained directly from a density functional theory (DFT) or Hartree Fock calculation, but they can also be further optimized with VMC\cite{marchimol}. It is well known that the antisymmetric product of these MO leads to the determinant of the matrix in which every molecular orbital is evaluated for each electron position:
\begin{equation}
  \Phi_{SD} (\mathbf{X}) = \det \left( 
	  \begin{matrix}
	  	\Phi^{mol}_1(\mathbf{r_1}) & \Phi^{mol}_1(\mathbf{r_2}) & \cdots & \Phi^{mol}_1(\mathbf{r_N}) \\
		\Phi^{mol}_2(\mathbf{r_1}) & \Phi^{mol}_2(\mathbf{r_2}) & \cdots & \Phi^{mol}_2(\mathbf{r_N}) \\
		\vdots & \vdots & \ddots & \vdots \\
		\Phi^{mol}_N(\mathbf{r_1}) & \Phi^{mol}_N(\mathbf{r_2}) & \cdots & \Phi^{mol}_N(\mathbf{r_N})   
  	\end{matrix}
  \right).
  \label{sd}
\end{equation}
For weakly correlated systems the JSD can often give reasonably good results with a reasonable computational cost and a limited number of variational parameters. It is also a common choice to use a linear combinations of SDs to improve the description of the WF, with {\it ansatzs} that take different names depending on the type and number of SDs considered. In this paper we will compare directly the results of our WFs to the  ones obtained with one of the most successful multi-determinant WFs, the full valence complete active space (FVCAS) WF.

\subsection{The Pairing Function}

The use of the pairing function in correlated wave  functions allows an electronic description 
that goes   beyond the single particle picture of the SD. The building block of this WFs has the following general form 
\begin{equation}
 	 f(\mathbf{r}_1 {\sigma_1},\mathbf{r}_2 {\sigma_2}) =\sum_{k,l} \lambda^{\sigma_1 \sigma_2}_{k,l} \phi_{k\sigma_1}(\mathbf{r_1}) \phi_{l\sigma_2}(\mathbf{r_2}),
 		\label{pair}
\end{equation}
where all the elements of the matrix $\lambda$  represent  most of the  wave function 
variational parameters.
They depend on the orbitals considered and on the spin $\sigma_1,\sigma_2$ of the so called geminal function $f$. In principle when we break the spin symmetry the basis sets 
used for $\uparrow$ and $\downarrow$ electrons can be different, otherwise the basis chosen does 
not  depend on the spin component. In 
order to set up a consistent many-body WF starting  from the geminal several choices are possible depending on the criteria adopted for the definition of  the geminal.  To highlight the different possibilities  we can recast Eq.~(\ref{pair}) in a way in which the spin dependency is more explicit

\begin{eqnarray}
 	 f(\mathbf{r}_1 {\sigma_1},\mathbf{r}_2 {\sigma_2}) &=& {1 \over \sqrt{2}} (|\uparrow \downarrow \rangle  -|\downarrow \uparrow \rangle ) f_+( \mathbf{r}_1,\mathbf{r}_2)  \nonumber \\
											 &+&  {1 \over \sqrt{2}} (|\uparrow \downarrow \rangle  +|\downarrow \uparrow \rangle )   f_-( \mathbf{ r}_1,\mathbf{ r}_2) \nonumber \\
											 &+& | \uparrow \uparrow \rangle f_{\uparrow} (\mathbf{r}_1,\mathbf{ r}_2) +  | \downarrow \downarrow \rangle f_{\downarrow} (\mathbf{r}_1, \mathbf{r}_2).
  \label{pair-spin}
\end{eqnarray}
where
\begin{eqnarray}
	f_\pm( \mathbf{r}_1,\mathbf{r}_2) &=&  f(\mathbf{r}_1 \uparrow,\mathbf{r}_2 \downarrow) \pm  f(\mathbf{r}_1 \downarrow,\mathbf{r}_2 \uparrow) \nonumber, \\
	f_{\sigma} (\mathbf{r}_1, \mathbf{r}_2)&=& f(\mathbf{r}_1 \sigma,\mathbf{r}_2 \sigma) \mbox{ with } \sigma= \uparrow,\downarrow .
  \label{explicit-terms}
\end{eqnarray}
In order to satisfy the Pauli principle we have $ f_\pm ( \mathbf{r}_1,\mathbf{r}_2)= \pm  f_{\pm}( \mathbf {r}_2,\mathbf{r}_1)  $
 and $f_\sigma( \mathbf r_1,\mathbf r_2)= -f_\sigma( \mathbf{ r}_2,\mathbf {r}_1) $ for $\sigma= \uparrow,\downarrow$.  
Our WF is then obtained by antisymmetrizing the product over all the electron pairs considered that, by definition, occupy  the {\em  same} pairing function. For simplicity we will enumerate the spin up electrons from $1$ to $N_\uparrow$ and the spin down ones from $N_\uparrow+1$ to $N$. 

As suggested by the name AGP, our goal is to define a WF that is literally the antisymmetrized product of the geminals and the unpaired orbitals (if present), namely 
\begin{eqnarray}
 	 \Phi(\mathbf{X}) =  \sum_\alpha  {\rm Sgn}(\alpha) \big( f(\mathbf{r}_{1_\alpha} {\sigma_{1_\alpha}},\mathbf{r}_{2_\alpha} {\sigma_{2_\alpha}})f(\mathbf{r}_{3_\alpha} {\sigma_{3_\alpha}},\mathbf{r}_{4_\alpha} {\sigma_{4_\alpha}}) \cdots  \nonumber \\
	   f(\mathbf{r}_{p-1_\alpha} {\sigma_{p-1_\alpha}},\mathbf{r}_{p_\alpha} {\sigma_{p_\alpha}})\Theta_1(\mathbf{r}_{p+1_\alpha})\cdots \Theta_{N-p}(\mathbf{r}_{N_\alpha})	 \big), \nonumber \\
	 \label{general-agp}
\end{eqnarray}
where $\alpha$ is one of the possible way of distributing the $N$ electrons between the $p/2$ pairs and the $N-p$ unpaired orbitals $\Theta$ and ${\rm Sgn}(\alpha)$ is the sign 
of the corresponding permutation of the particles that is  required to insure the fermionic behaviour. In particular different choices  of the pairing function, 
obtained by  excluding  one or more  terms in the Eq.~(\ref{explicit-terms}), lead to different ways to compute Eq.~(\ref{general-agp}). These choices also  impact quantitatively and qualitatively on the kind of physics that we can describe by means of this type of  WF. Therefore, we will distinguish in the following among three distinct cases: if we consider only the singlet term in Eq.~(\ref{explicit-terms}) we  obtain the AGPs,  if we include the singlet and the $S_z=0$ triplet term we have the AGPu, while the most general case is just  the definition adopted here for the AGP.

\subsubsection{AGPs}

Let us consider for the moment the unpolarized case $N_\uparrow=N_\downarrow$, the extension to the polarized cases will be straightforward and will be discussed later on.
When no triplet correlations are allowed we build our WFs using only singlet pairs and the pairing function in Eq.~(\ref{pair-spin}) contains only the symmetric element $f_+$ 
\begin{equation}
 	 f(\mathbf{r}_1 {\sigma_1},\mathbf{r}_2 {\sigma_2}) = {1 \over \sqrt{2}} (|\uparrow \downarrow \rangle  -|\downarrow \uparrow \rangle ) f_+( \mathbf{r}_1,\mathbf{r}_2).
  \label{agps-pair}
\end{equation}
In this case we project a perfect singlet  that we  denote as AGPs. The  $\lambda$ matrix elements 
in Eq.~(\ref{agps-pair}) 
are non zero only for $\sigma_1 \ne \sigma_2$ and they are  symmetric for spin exchange. 
In order to calculate the AGPs we can write all the possible combinations of pairs of opposite spin electrons 
in a matrix defined as
\begin{equation}
  F= \left(
  \begin{matrix}
	  f(\mathbf{r}_1 \uparrow,\mathbf{r}_{N_\uparrow+1} \downarrow) &  f(\mathbf{r}_1 \uparrow,\mathbf{r}_{N_\uparrow+2} \downarrow)  & \cdots &  f(\mathbf{r}_1 \uparrow,\mathbf{r}_{N} \downarrow) \\
	  	  f(\mathbf{r}_2 \uparrow,\mathbf{r}_{N_\uparrow+1} \downarrow) &  f(\mathbf{r}_2 \uparrow,\mathbf{r}_{N_\uparrow+2} \downarrow)  & \cdots &  f(\mathbf{r}_2 \uparrow,\mathbf{r}_{N} \downarrow) \\ 
		  \vdots & \vdots & \ddots & \vdots \\
		  f(\mathbf{r}_{N_{\uparrow}} \uparrow,\mathbf{r}_{N_\uparrow+1} \downarrow) &  f(\mathbf{r}_{N_{\uparrow}} \uparrow,\mathbf{r}_{N_\uparrow+2} \downarrow)  & \cdots &  f(\mathbf{r}_{N_{\uparrow}} \uparrow,\mathbf{r}_{N} \downarrow) 
   \end{matrix}
   \right).
  \label{agps_matrix}
\end{equation}
In this way to each row of the matrix  corresponds an electron of spin $\uparrow$, and to each column an electron of spin $\downarrow$. The definition of the matrix $F$ in this form is convenient  because it allows the antisymmetrization requested by the Eq.~(\ref{general-agp}) in a simple and efficient way. Indeed, it can be demonstrated\cite{doi:10.1063/1.1604379} that the correct antisymmetrization of the pairs considered in this case is given by 
\begin{equation}
  \Phi_{AGPs}(\mathbf{X})=\det {F}.
  \label{agps}
\end{equation}
This is somehow intuitive, because we want to sum all the possible products of $N/2$ matrix elements of $F$,  where  in all these factors a column element or  a row element is present only once, exhausting all the possible configurations of the system considered with an appropriate  
$\pm$ sign that, in this case is just given by the  one corresponding to the 
determinant of $F$.

When the system is polarized and $N_\uparrow \ne N_\downarrow$, we cannot build the solution using only the singlet terms, because the  matrix $F$ written as  in Eq.~(\ref{agps_matrix}) is a rectangular matrix and its determinant cannot be  computed. 
Supposing for simplicity that $N_\uparrow > N_\downarrow$, in  this case  we have  to  add a number
$N_\uparrow - N_\downarrow$
of unpaired  spin-up MOs $\{ \Theta_i(\mathbf{r}) \}$ not only for  fulfilling
the  polarization required but, most importantly, to turn  the matrix $F$ 
to a perfectly defined square matrix: 
\begin{equation}
  F= \left(
  \begin{matrix}
	  f(\mathbf{r}_1 \uparrow,\mathbf{r}_{N_\uparrow+1} \downarrow) &  \cdots  & \Theta_1(\mathbf{r}_1) &  \cdots & \Theta_{N_\uparrow-N_\downarrow}(\mathbf{r}_1) \\
	  f(\mathbf{r}_2 \uparrow,\mathbf{r}_{N_\uparrow+1} \downarrow) &  \cdots  & \Theta_1(\mathbf{r}_2) &  \cdots & \Theta_{N_\uparrow-N_\downarrow}(\mathbf{r}_2) \\
	  \vdots & \ddots & \vdots & \ddots & \vdots \\
	  f(\mathbf{r}_{N_\uparrow} \uparrow,\mathbf{r}_{N_\uparrow+1} \downarrow) &  \cdots  & \Theta_1(\mathbf{r}_{N_\uparrow}) &  \cdots&  \Theta_{N_\uparrow-N_\downarrow}(\mathbf{r}_{N_\uparrow}) .
	    \end{matrix}
   \right)			  
  \label{agps_unpaired}
\end{equation}

Also in this case a consistent antisymmetric wave function can be again calculated as the determinant\cite{doi:10.1063/1.1604379} of the matrix $F$ exactly in the  same way  of the singlet pairing   in Eq.~(\ref{agps}).

\subsubsection {AGPu}

For the AGPu only the parallel  spin term of  the triplet component are omitted. This means that the spin symmetry is broken and a magnetic order parameter can be directed along the $z-$quantization axis. This WF is called broken symmetry AGP (AGPu) and the difference from the previous AGPs is the presence of the antisymmetric $f_-$ component  in the definition of the pairing function in Eq.~(\ref{pair-spin}), that for this case is
\begin{eqnarray}
 	 f(\mathbf{r}_1 {\sigma_1},\mathbf{r}_2 {\sigma_2}) &=& {1 \over \sqrt{2}} (|\uparrow \downarrow \rangle  -|\downarrow \uparrow \rangle ) f_+( \mathbf{r}_1,\mathbf{r}_2)  \nonumber \\
											 &+&  {1 \over \sqrt{2}} (|\uparrow \downarrow \rangle  +|\downarrow \uparrow \rangle )   f_-( \mathbf{ r}_1,\mathbf{ r}_2).
	  \label{pair-agpu}
\end{eqnarray}

In order to define this pairing function we break the spin symmetry in the opposite  electron  
spin case with $\sigma_1 \ne \sigma_2$, by  keeping equal to zero the $\sigma_1 = \sigma_2$ components of Eq.~(\ref{pair}). With exactly the same procedure used in the case of the AGPs, depending on the polarization, we can build the same matrix $F$ of Eq.~(\ref{agps_matrix}) or Eq.~(\ref{agps_unpaired}), that now is no longer symmetric. Even in this case the correct antisymmetrized sum of these pairs is given by the determinant\cite{doi:10.1063/1.1604379}. Thus, analogously to Eq.(\ref{agps}) we obtain
\begin{equation}
  \Phi_{AGPu}(\mathbf{X})=\det {F},
  \label{agpu}
\end{equation}
that implements the simplest broken symmetry {\it ansatz }based on the pairing function.

\subsubsection{AGP}

The AGP (also known in literature as Pfaffian WF\cite{PhysRevLett.96.130201}) is in our opinion the most important pairing function, being the most general one and encoding new variational freedoms into the AGPs and the AGPu. We will show that it represents the most powerful description of the chemical bond within the paradigm developed in this work. This WF represents also  the most general mean-field state, namely the GS of a mean-field Hamiltonian containing also BCS anomalous terms projected on a given number $N$ of particles and total spin projection $S^z_{tot}=\sum\limits_{i=1}\sigma_i$ along the $z-$ quantization axis. In this case the definition of the pairing function is exactly the one in Eq.~(\ref{pair-spin}), containing all the terms including the parallel spin terms of the triplet. This means that now, when we build the AGP, we have to include in the WF also the parallel spins electron pairs. In this way the AGP can also describe a  magnetic order parameter in any direction of the space, and thus it is also possible to rotate the spin component of the WF in any direction. This will allow us to break the symmetry along the spin quantization axis and then rotate it. As we will explain later, this plays a crucial role when we use this WF in combination with our JF, since it allows us to preserve the total $S_z$ of the molecules and include spin fluctuations.

Of course we cannot create a WF using only pairs if the number of electrons in the system is odd, so, for the moment, let us assume $N$ is even. The extension to the odd number of electrons is trivial and will be discussed immediately after. We will dub as $W$ the matrix containing all the possible pairs
\begin{equation}
  W= \left(
  \begin{matrix}
  	0 & f(\mathbf{r}_1 \uparrow, \mathbf{r}_2 \uparrow) & \cdots & f(\mathbf{r}_1 \uparrow, \mathbf{r}_N \downarrow ) \\
  	f(\mathbf{r}_2 \uparrow, \mathbf{r}_1 \uparrow) & 0 & \cdots & f(\mathbf{r}_2 \uparrow, \mathbf{r}_N \downarrow ) \\
	\vdots & \vdots & \ddots & \vdots \\
	f(\mathbf{r}_{N-1} \downarrow, \mathbf{r}_1 \uparrow) & f(\mathbf{r}_{N-1} \downarrow, \mathbf{r}_2 \uparrow) & \cdots & f(\mathbf{r}_{N-1} \downarrow, \mathbf{r}_N \downarrow ) \\
		f(\mathbf{r}_N \downarrow, \mathbf{r}_1 \uparrow) & f(\mathbf{r}_N \downarrow, \mathbf{r}_2 \uparrow) & \cdots & 0
  \end{matrix}
  \right),
  \label{pfaff_matrix}
\end{equation}
where the matrix is antisymmetric for the fermionic commutation rules and thus the elements of the diagonal are set to zero.
We can recast the $W$ highlighting its different spin sectors as
\begin{equation}
  W= \left(
  \begin{matrix}
	W_{\uparrow \uparrow} & W_{\uparrow \downarrow} \\
        W_{\downarrow \uparrow} & W_{\downarrow \downarrow} \\
  \end{matrix}
  \right)
  \label{pfaff_matrix_small}
\end{equation}
where $W_{\uparrow \uparrow}$ and $W_{\downarrow \downarrow}$ are respectively a $N_\uparrow \times N_\uparrow $ and a $N_\downarrow \times N_\downarrow $  
antisymmetric matrices that take into account the parallel spin terms of the triplet, while $W_{\uparrow \downarrow}$ is a $N_\uparrow \times N_\downarrow $ matrix such 
that $W_{\uparrow \downarrow} = -W^T_{\downarrow \uparrow}$ describing the remaining triplet and singlet contributions. In the case of AGPs and AGPu we can also build a similar matrix where the matrices $W_{\uparrow \uparrow}$ and $W_{\downarrow \downarrow}$ are identically zero.

Analogously to the case of the AGPs and AGPu, we have to identify a way to calculate the antisymmetric product of all the pairs considered. In this case it is easy to identify the antisymmetrization procedure defined in Eq.~(\ref{general-agp}) as the Pfaffian of the matrix $W$. After introducing this algebraic operation the reason will be straightforward to the reader. 

The Pfaffian is an algebraic operation acting on antisymmetric square matrices with an even number of rows and columns. Being $N$ even, the matrix $W$ satisfies these hypothesis. The usual definition of the Pfaffian, requires the introduction of the concept of partition of the matrix $W$
\begin{equation}
  A(\alpha)=sign(\alpha)\prod_{k=1}^{N/2} W_{i_k,j_k}
  \label{partition}
\end{equation}
where all $i_k$ and $j_k$ are different, $i_k<j_k$ for each $k$ and $i_1<i_2<\dots<i_N$. The $sign(\alpha)$ is given by the permutation that orders the vector of the indices $\{i_1,j_1,i_2,j_2,\dots,i_M,j_M\}$. In this way all the indices are considered only once. The Pfaffian is then defined as
\begin{equation}
   \text{Pf}(W)=\sum_\alpha A(\alpha)
  \label{pf}
\end{equation}
where the sum over $\alpha$ is extended over all the possible partitions. However an alternative definition\cite{doi:10.1063/1.1703953} of the Pfaffian can better clarify the correspondence to the Eq.~(\ref{general-agp}). It can indeed be defined alternatively as
\begin{equation}
	\text{Pf}(W)= \left[ (N/2)! 2^{N/2} \right]^{-1} \sum_P \text{sign}(P) \prod_{k_P=1}^{N/2} W_{i_{k_P},j_{k_P}}
	\label{kasteleyn}
\end{equation}
where $P$ now represents a generic permutation of the possible row and column indices of the matrix without any constraints and the $sign(P)$ is the parity of the permutation. In this definition it is easy to recognize the antisymmetrized sum corresponding to the Eq.~(\ref{general-agp}). Let us introduce now a further property of the Pfaffian that will be useful in the following. In the following 
we will indicate with  $0$ a $m\times m$ matrix containing only $0$ and B a generic $m\times m$ matrix,  we have that
\begin{equation}
   \text{Pf}\left[
   	 \begin{matrix}
   		0 & B\\
		-B^T & 0
   	\end{matrix}\right]=(-1)^{m(m-1)/2} \det(B).
  \label{pftoagp}
\end{equation}

For odd number of electrons it is necessary to use a spin-dependent  unpaired orbital $\Theta_\sigma(\mathbf{r})$ so that we can accommodate the remaining electron that is not considered by the product of the pairs.  The unpaired orbital introduces a supplementary row and column to the matrix $W$. Being $\Theta_\uparrow=(\Theta_\uparrow(\mathbf{r}_1),\Theta_\uparrow(\mathbf{r}_2), \cdots, \Theta_\uparrow(\mathbf{r}_{N_{\uparrow}}))$ the vector containing the values of the unpaired orbital $\Theta_\uparrow$ at the $\uparrow$ electron positions and $\Theta_\downarrow=(\Theta_\downarrow(\mathbf{r}_{N_\uparrow+1}),\Theta_\downarrow(\mathbf{r}_{N_\uparrow+2}), \cdots, \Theta_\downarrow(\mathbf{r}_N))$ the one calculated for the $\downarrow$ electron ones, we modify the matrix in Eq.~(\ref{pfaff_matrix_small}) as
\begin{equation}
  W= \left(
  \begin{matrix}
	W_{\uparrow \uparrow} & W_{\uparrow \downarrow} & \Theta_\uparrow \\
        W_{\downarrow \uparrow} & W_{\downarrow \downarrow} &  \Theta_\downarrow \\
        -\Theta^T_\uparrow & -\Theta^T_\downarrow & 0
  \end{matrix}
  \right).
  \label{pfaff_matrix_small2}
\end{equation}
Also in this case the permutation sum  implied  by the Pfaffian leads to  the correct antisymmetrization required from Eq.~(\ref{general-agp}). The matrix W satisfies the hypothesis of the calculation having an even leading matrix dimension $\bar{N}=N+1$. We can further notice that no assumption has been made on the polarization of the system and so no unpaired orbital is required except for a single one in case of odd $N$.

It is however possible in principle to introduce further pairs of unpaired orbitals, if, for example, we want to describe an AGPs or AGPu with a full AGP WF. We define $\Theta_{i\sigma}(\mathbf{r}) $ as the set of the considered $m$ unpaired orbitals orbitals and $\Theta_{i\uparrow}=(\Theta_{i,\uparrow}(\mathbf{r}_1),\Theta_{i,\uparrow}(\mathbf{r}_2), \cdots, \Theta_{i,\uparrow}(\mathbf{r}_{N_{\uparrow}}))$ the vector containing the values of the unpaired orbital $\Theta_{i,\uparrow}$ for the $\uparrow$ electron positions and $\Theta_{i\downarrow}=(\Theta_{i,\downarrow}(\mathbf{r}_{N_\uparrow+1}),\Theta_{i,\downarrow}(\mathbf{r}_{N_\uparrow+2}), \cdots, \Theta_{i,\downarrow}(\mathbf{r}_N))$ the one calculated for the $\downarrow$ electron ones. We can modify the matrix in Eq.~(\ref{pfaff_matrix_small}) as
\begin{equation}
  W= \left(
  \begin{matrix}
	W_{\uparrow \uparrow} & W_{\uparrow \downarrow} & \Theta_{1\uparrow} & \cdots & \Theta_{m\uparrow} \\
        W_{\downarrow \uparrow} & W_{\downarrow \downarrow} &  \Theta_{1\downarrow} & \cdots & \Theta_{m\downarrow}  \\
        -\Theta^T_{1\uparrow} & -\Theta^T_{1\downarrow} & 0 &\ddots & \vdots\\
	\vdots & \vdots & \vdots & \ddots & \vdots \\
         -\Theta^T_{m\uparrow} & -\Theta^T_{m\downarrow} & 0 &\cdots & 0\\
  \end{matrix}
  \right),
  \label{pfaff_matrix_small3}
\end{equation}
that is a $\bar{N}\times \bar{N}$ matrix where $\bar{N}=N+m$. We can again antisymmetrize this product using the definition of the Pfaffian provided in Eq.~(\ref{kasteleyn}). A careful reader could have noticed that, by  applying the Pfaffian definition, we are antisymmetrizing not only over the electron indices but also over the orbital indices of the unpaired orbitals. This antisymmetrization, however, contains the one over the physical electrons, and leads therefore to a physically allowed electronic wave function.

Moreover we can notice that, by  using the previous definition, we can identify the AGPs and the AGPu as sub-cases of the general AGP. Indeed, by using the expressions  of the pairing function and the unpaired orbitals of the AGPs and AGPu we obtain $W_{\uparrow \uparrow}=0$, $W_{\downarrow \downarrow}=0$, $\Theta_{i\downarrow}=0$ and $\bar{N}=2N_\uparrow$. By merging Eq.~(\ref{agps_unpaired}) and Eq.~(\ref{pfaff_matrix_small}) we can define
\begin{equation}
 W= \left(
   	 \begin{matrix}
   		0 & F\\
		-F^T & 0
   	\end{matrix}\right),
  \label{pftoagp2}
\end{equation}
and this means that applying Eq.~(\ref{pftoagp}) we immediately obtain
\begin{equation}
   \text{Pf } (W)=\pm \det(F),
  \label{pftoagp3}
\end{equation}
where the sign only depends on the number of electrons and is constant, thus irrelevant. This shows in a straightforward way that the AGPs and AGPu defined in the previous subsection are nothing but particular cases of the most general AGP.

\subsection{Jastrow Factor}\label{sec:jf}

Within QMC, it is easy  to improve the quality of the WF by  multiplying the WF with an exponential JF. This last one enriches the description of the GS by encoding explicitly the electronic correlation, while speeding up the  convergence to the complete basis set limit\cite{PhysRevX.7.031059}. Indeed, with an appropriate choice, the JF can satisfy exactly  the electron-electron and electron-ion cusp conditions of the many-body WF, consequences  of the Coulomb $1/r$ singularity at short distance. In this paper we introduce a new kind of JF that contains a richer dependence on the spin and that plays a fundamental role when used in combination with the AGP WF. The JF is defined as
\begin{equation}
  J(\mathbf{X})=e^{U_{ei}+U_{ee}},
  \label{JF}
\end{equation}
where $U_{ei}$ is a single body term that deals explicitly with the electron-ion interaction and  $U_{ee}$ is a many-body term that properly accounts for the electronic correlation. The single body term is 
\begin{equation}
  U_{ei}=\sum_{i=1}^{N}u_{ei}(\mathbf{r}_i),
  \label{sb1}
\end{equation}
with $u_{ei}$ being
\begin{equation}
  u_{ei}(\mathbf{r}_i)= \sum_{I=1}^{\#ions} -Z_I  \frac{1-\exp(b_{ei}|\mathbf{r}_i-\mathbf{R}_I|)}{b_{ei}} + g_I(\mathbf{r}_i).
  \label{sb2}
\end{equation}
In Eq.~(\ref{sb2}) $Z_I$ is the atomic number of the atom $I$ and $b_{ei}$ is a variational parameter, while $g_I(\mathbf{r}_i)$ encodes 
the  most  general non homogeneous electron-ion one-body term, i.e. depending  explicitly on  all nuclear and electron coordinates and not  only on their relative distances, that is defined as
\begin{equation}
  g_{I}(\mathbf{r}_i)= \sum_{\nu} \xi_{I,\nu} \phi_{I,\nu}(\mathbf{r}_i),
  \label{sb3}
\end{equation}
where the summation is extended over all the gaussian orbitals in the JF basis set centered on the $I$-th atom.  
The electron-electron term instead is written as
\begin{equation}
  U_{ee}=\sum_{i<j}u_{ee}(\mathbf{r}_i \sigma_i, \mathbf{r}_j \sigma_j),
  \label{mb1}
\end{equation}
where the sum is extended over the pairs of different electrons and where
\begin{equation}
  u_{ee}(\mathbf{r}_i \sigma_i, \mathbf{r}_j \sigma_j)=k_{\sigma_i,\sigma_j}\frac{|\mathbf{r}_i-\mathbf{r}_j|}{1+b^{ee}_{\sigma_i,\sigma_j}|\mathbf{r}_i-\mathbf{r}_j|}
+g_{ee}(\mathbf{r}_i \sigma_i, \mathbf{r}_j \sigma_j),
  \label{mb2}
\end{equation}
with the $2\times 2$ matrix $b^{ee}_{\sigma,\sigma^\prime}$  described by 
one $b^{ee}_{\sigma,\sigma^\prime}=b^{ee}$ 
or two variational parameters for  $\sigma_i=\sigma_j$  
when  $k_{\sigma_i,\sigma_j}=1/4$ and $b^{ee}_{\sigma,\sigma^\prime}=b^{ee}_\parallel$ 
and for $\sigma_i \ne \sigma_j$ when 
$k_{\sigma_i,\sigma_j}=1/2$ and $b^{ee}_{\sigma,\sigma^\prime}=b^{ee}_\perp$.
The conventional expression for the JF can be obtained by removing all spin depencency 
in  the previous expressions and remaining  only with the variational parameters 
corresponding to the opposite  spin case $k_{\sigma_i,\sigma_j}=1/2$ and  $b^{ee}_{\sigma,\sigma^\prime}=b^{ee}$.

In our expression the first term in Eq.~(\ref{mb2}) deals explicitly with the electron-electron cusp conditions, the second term in Eq.~(\ref{mb2})  instead is a bosonic pairing function in the form
\begin{equation}
 	 g_{ee}(\mathbf{r}_1 {\sigma_1},\mathbf{r}_2 {\sigma_2}) =\sum_{k,l} \zeta^{\sigma_1 \sigma_2}_{k,l} \phi_{k\sigma_1}(\mathbf{r_1}) \phi_{l\sigma_2}(\mathbf{r_2}),
 		\label{pair-jf}
\end{equation}
with the  elements of the matrix  $\zeta$  defining further variational parameters.   
Notice that both $g_I$ and $g_{ee}$ do not affect the  cusp conditions 
because they are expanded  over cuspless gaussian  orbitals. The $g_{ee}$ term has the same form of Eq.~(\ref{pair}), but, since the fermionic behaviour is already encoded in the fermionic part of the WF, this term is symmetric under particle exchange. The use of a pairing function in the JF enriches the description of the charge and spin correlations of the system noticeably improving the quality of the global WF. It is a common practice to adopt a simplified or even absent spin dependency in the function $u$ of Eq.~(\ref{mb2}). This is often accurate  for systems where the magnetic properties are not relevant. We will refer to it in the following with the prefix Js in the WF, in contrast with the prefix J used for the full spin dependent JF. 

A perfect singlet remains such after the multiplication of a spin independent Jastrow, and so our spin dependent JF is not appropriate  if we do not want to break the spin symmetry.
 It is, instead,  necessary if we want to recover, at  least approximately,  the singlet from a spin contaminated  broken symmetry ansatz. A general spin dependent $u$, as defined in Eq.~(\ref{mb2}), 
 is therefore of fundamental importance for  the AGPu or the AGP  ansatzs. 

 Let  us start  with a simple example. We consider two atoms with opposite spins, and 
 break the spin symmetry by   orienting   the spins of the atoms 
 along the $z-$quantization axis.
In  this case
 the JF is not able to change  the  classical antiferromagnetic spin state because it acts as an irrelevant constant when applied to it. It is instead more physical to orient the spin moment of the atoms in a direction perpendicular to the quantization axis chosen for the JF. In this way the JF can act on the electrons and the spins while the magnetic moment is free to fluctuate and recover its genuine quantum character. As previously mentioned with the AGP it is possible to rotate the spin of the WF in every direction and orient the magnetic moment in  any direction  of  the  space. This works particularly well in combination with our Jastrow that can suppress the unfavored triplet configurations with parallel  spins generated by the rotation.  This optimal spin-orientation of the atoms, i.e.  perpendicular to the JF one,  is rigorously valid within the well known spin-wave theory of a quantum antiferromagnet\cite{10.1143/PTP.97.399}. In this case the JF defined with a spin-quantization axis perpendicular to the magnetic moment of the atoms allows the description of  the quantum fluctuations and the corresponding zero point energy, even for a finite (as is our case) number of atoms\cite{10.1143/PTP.97.399}. 

\subsection{Procedure}

The first step to calculate and optimize our WF is to identify a reasonable starting point. We chose to start from a DFT calculation because of its flexibility.  We have used LDA calculations for spin symmetric systems, while for the ones with opposite spin antiferromagnetic moments we have broken the symmetry with a LSDA calculation, allowing  an accurate initialization of the WF. The SD obtained from a DFT calculation is mapped without loss of information into an AGPs or AGPu and then in a second analogous step we convert the AGPs and AGPu into a full AGP.

For the first conversion let us consider Eq.~(\ref{pair}). If we compute it in the basis set of the MOs obtained from the DFT we have
\begin{equation}
 	 f(\mathbf{r}_1 \uparrow, \mathbf{r}_2 \downarrow) =\sum_{\alpha} \bar{\lambda}^{\uparrow \downarrow}_{\alpha,\alpha} \Phi^{mol}_{\alpha\uparrow}(\mathbf{r_1}) \Phi^{mol}_{\alpha\downarrow}(\mathbf{r_2}),
 		\label{init-agp}
\end{equation}
namely  only the diagonal terms in the matrix $\bar{\lambda}$ are present. Moreover we can also remove the spin dependence if there is no symmetry breaking. For the polarized case the unpaired orbitals are the last occupied MOs. 
By  substituting the definition of the 
molecular orbitals with their expansion over a localized atomic basis set, 
as given in Eq.~(\ref{molorb}),   we can recast the above equation exactly in the  same 
form shown in Eq.(\ref{pair}) with a 
matrix $\lambda$:
\begin{equation}
 	\lambda_{k,l} =\sum_{k} P^\dagger_{k,\alpha}\bar{\lambda}^{\uparrow \downarrow}_{\alpha,\alpha} P_{\alpha,l}.
 		\label{init-agp3}
\end{equation}

When we convert an AGPs or AGPu into an AGP WF we already have an initialization for the sectors of the pairs with different spins $W_{\uparrow \downarrow}$ and $W_{\downarrow \uparrow}$ that can be obtained directly from the AGPs or AGPu pairing functions.  The main challenge is to find a reasonable initialization for the two sectors $W_{\downarrow \downarrow}$ and $W_{\uparrow \uparrow}$ that are not described by the AGPs or AGPu. 

There are two different procedures that we can follow, the first one is used for polarized systems, the second one instead is preferred in case of broken spin symmetry and in presence of antiferromagnetism, namely molecules well described  by opposite atomic magnetic moments. If there is no antiferromagnetism and the polarization is such that $|S^z_{tot}|<1$ the  $W_{\downarrow \downarrow}$ and $W_{\uparrow \uparrow}$ are instead identically zero. This holds 
not only for $S^z_{tot}=0$ but also for $S^z_{tot}=\pm 1/2$, where the single unpaired molecular  
orbital used in Eq.~(\ref{pfaff_matrix_small2}), acquires also a  spin  dependency, not present 
in the AGPs and  AGPu cases.

Obviously the atoms, but also the O${}_2$ molecule, do not have antiferromagnetism, but, on the other side, they have a net polarization. We can build the $W_{\uparrow \uparrow}$ block of the matrix using the two unpaired orbitals $\Theta_1$ and $\Theta_2$ 
for the definition  of the parallel spin matrices of the AGPs or AGPu in the following way
\begin{equation}
 	 f(\mathbf{r}_1 \uparrow, \mathbf{r}_2 \uparrow) = \bar{\lambda} (\Theta_{1}(\mathbf{r_1}) \Theta_{2}(\mathbf{r_2}) - \Theta_{1}(\mathbf{r_2}) \Theta_{2}(\mathbf{r_1})),
 		\label{init-upup}
\end{equation}
here the presence of the minus sign guarantees the pairing function to be antisymmetric under particle exchange, while the $\bar{\lambda}$ is an arbitrary scaling factor that has no influence on the final value of the WF. Once we map the unpaired orbitals in the desired basis set we obtain the variational parameters of the matrix $\lambda$ for the $\uparrow \uparrow$ sector.

In presence of opposite atomic magnetic moments,  it is possible to rotate the spin component of the pairing function to initialize the $W_{\downarrow \downarrow}$ and $W_{\uparrow \uparrow}$ sectors. As we mentioned earlier a further effect of this operation is to direct the atomic magnetic moments in a direction perpendicular to the spin quantization axis. It is worth mentioning that, within our method, the spin orientation with respect to the molecular axis is irrelevant since in a non relativistic Hamiltonian  the spin-orbit coupling is not present. In this case we have chosen to work with  the atomic magnetic moments  perpendicular to the $z-$axis, hence we applied a rotation of $\pi/2$ around the $\hat{y}$ direction. This operation maps
\begin{equation}
	|\uparrow \rangle \rightarrow \frac{1} {\sqrt{2}} \left( |  \uparrow \rangle + |\downarrow \rangle \right)   \mbox{ and }  |\downarrow  \rangle  \rightarrow  \frac 1 {\sqrt{2}} \left( | \uparrow  \rangle - |\downarrow \rangle \right) .	
	  \label{rot-ud}
\end{equation}
If we apply this transformation to the pairing function from Eq.~(\ref{pair-agpu}) we obtain 
\begin{eqnarray}
 	 f(\mathbf{r}_1 {\sigma_1},\mathbf{r}_2 {\sigma_2}) &=& {1 \over \sqrt{2}} (|\uparrow \downarrow \rangle  -|\downarrow \uparrow \rangle ) f_+( \mathbf{r}_1,\mathbf{r}_2)  \nonumber \\
									 &+& | \uparrow \uparrow \rangle  f_-( \mathbf{ r}_1,\mathbf{ r}_2)  -  | \downarrow \downarrow \rangle  f_-( \mathbf{ r}_1,\mathbf{ r}_2).
  \label{init-agp2}
\end{eqnarray}

This transformation provides a meaningful initialization to our AGP WF that now have to be optimized to reach the best possible description of the GS. Indeed, within VMC, it is only thanks to the optimization that we can improve the description of the GS. So far we have only converted the DFT WF from one {\it ansatz }to another, but the key for the success of this procedure is the optimization of all the possible variational parameters. It is indeed crucial to optimize not only the ones corresponding to the matrix $\lambda$ and the JF parameters, but also the coefficients of the hybrid orbitals $\mu$ and the exponents of the gaussian basis set $Z_\nu$. This is realized computationally in a very efficient way using a coding technique called Adjoint Algorithmic Differentiation\cite{doi:10.1063/1.3516208} that allows calculations of  total energy derivatives with respect to all the variational parameters involved in a given algorithm that computes only the energy. This is remarkably done by paying a very small slowing down of a factor $\approx 2-3$  with respect to the latter algorithm. We have also used a state of the art optimization scheme\cite{2007SOR,2007UMR}  for a correct search of the energy minimum.  Remarkably,  even when there is some possible  dependency among the many variational parameters considered in our ansatz,  the Stochastic Reconfiguration technique remains stable and  efficient, thanks   to an appropriate regularization of the of the stochastic matrix $S$\cite{becca_sorella_2017}. Once we calculate the variational minimum, the best description of the GS is then obtained with the DMC calculation.

 Even considering that  the number of variational parameters involved in the calculation may be quite large, the optimization has a very small impact over the total computational cost, that is indeed mostly  given by the DMC for all the cases  reported in this work. In Tab. \ref{comp_costs} we compare the computational cost of the DMC calculations for the different WFs considered. We notice that the JAGP is even less expensive than the JSD and JsAGPs WFs. The JAGP and JAGPu are so efficient because,  in this case, the variance of the energy is considerably  smaller as we can see from Table \ref{comp_costs}. This impies that JAGP and JAGPu  require a smaller number of DMC iterations to reach the desired accuracy,  because they have lower variance compared with the JsAGPs and the JSD, thanks to the spin dependent JF.

For large number $N$ of electrons,  the DMC calculation should scale as $N^4$ for fixed  total energy accuracy, and  the main question, that we have not  studied  here, is whether the optimization remains  computationally  negligible  because the number of variational parameters  scales as $N^2$. In  this respect we have experienced that an optimization technique performed with a slow but  very stable method,  namely with a large number of ''cheap''  optimization steps,  each one determined by  a relatively small number of samples (even much smaller than the number of parameters) is very promising for future large $N$ applications.

\begin{table}
  \caption{DMC computational time required to obtain an accuracy of $0.1$ mH and energy variance
  on the oxygen dimer with an Intel Xeon architecture using a recent  LRDMC  algorithm\cite{PhysRevB.101.155106} with a  lattice spacing equal to $0.05\ Bohr$, the smallest used  in this work. In these systems the cost for doing about 10000 iterations for the  VMC 
  optimization of our WFs is less than 30hours. The CPU time reported in the table  corresponds to the total one
(time spent by a  single  core times the number  of cores) for obtaining
the  required accuracy, e.g.  with   256 cores parallel computation
the  JAGP calculation can be  obtained with about seven hours of walltime. 
  } \label{comp_costs}
  \begin{tabular}{l c c}
    \hline
    \hline
    \rule[-0.4mm]{0mm}{0.4cm}
    WF &   CPU time & Variance[$H^2$] \\
    \hline
    \rule[-0.4mm]{0mm}{0.4cm}
   JSD &   2806	& 	2.909 \\
\rule[-0.4mm]{0mm}{0.4cm}
JsAGPs &	2526	& 	2.819 \\
\rule[-0.4mm]{0mm}{0.4cm}
JAGPu & 14523	& 2.455\\
\rule[-0.4mm]{0mm}{0.4cm}
   JAGP& 1857.79	& 2.125  \\
   \hline
    \hline
\end{tabular}                   
\end{table}

Finally we introduced a technique to deal with particularly unstable AGP WFs. Indeed, it is possible, after a very large number of optimization steps ($>10000$), that some eigenvalues of the matrix 
$\lambda$  become too small as compared with the largest reference  eigenvalue. This creates some instabilities in the inversion of the matrix $W$, required for the QMC fast updates. For this reason, by  an appropriate  
use of the PFAPACK library\cite{Wimmer:2012:A9E:2331130.2331138}, we identified a procedure to map the diagonalization of a full skew-symmetric matrix $\lambda$ to the one corresponding  to a real tridiagonal symmetric matrix. After this mapping we can use the most powerful and stable LAPACK routines for diagonalization. Indeed, most linear algebra packages  cannot deal with antisymmetric matrices and a general diagonalization tool  was not available for this case. The introduction of this procedure, described in details in the appendix, 
allows us to describe the matrix $\lambda$ in terms of eigenvalues and orthogonal orbitals playing the role  of  eigenvectors of an antisymmetric matrix. We will refer to them in the appendix as molecular orbitals since it may  represent a rather  formal definition of them, within  the formulation introduced  in  this work. With this meaningful decomposition we can finally regularize the matrix $\lambda$ by replacing the too small eigenvalues with reasonable lower bounds and continue, if necessary, with the optimization of the variational parameters. 

\subsection{The $S^2$ operator}

The basic concept of QMC relies on the real space configurations sampling of a general 
electronic system. All the observables can be indeed calculated in the basis 
where the electron positions and their spins are defined. In particular, for the systems considered it is interesting to estimate the spin observables in order  to understand their magnetic properties and the quality of the corresponding WFs. If during the simulation the value of $S_z$ is fixed, when we break the symmetry the value of the $S^2$ is instead the result of the interplay between the JF and the AGP or the AGPu. The efficient computation of the expectation value of the $S^2$ operator has already been described in \cite{doi:10.1021/ct401008s} for the JsAGPu and will be shown now for the JAGP. 

In the following we will show how to evaluate $S^2$ in a region of the space with a fast and computationally cheap approach based on the fast update algebra of the AGP and the spin dependent JF. Let us consider the expectation value of the $S^2$ operator over a generic WF $\Psi$ by direct application of its definition.  In the following we use the completeness of the spatial configurations:
\begin{equation}
\sum\limits_\mathbf{X} | \mathbf{X} \rangle \langle \mathbf{X} | = {\bm I} 
\end{equation}
where  the  summation symbol imply here  also a $3N-$multidimensional integral over the electron coordinates.
Assuming a fixed polarization we can write the explicit expression of the total spin square as
\begin{eqnarray}
	\langle S^2 \rangle_\Psi & = & 
 \frac{\langle \psi \vert (S^z)^2 \vert \psi \rangle}{\langle \psi \vert \psi \rangle}	+{1 \over 2} \sum\limits_\mathbf{X} \sum_{i,j}^N 	\frac{\langle \psi \vert (S_i^+ S_j^- +  S^-_i S^+_j) \vert \mathbf{X} \rangle \langle \mathbf{X} \vert \psi \rangle}{ \langle \psi \vert \psi \rangle }
		\nonumber \\ 	
	 & = & \sum_{i=1}^{N_{\uparrow}} \sum_{j=N_{\uparrow}+1}^N 
	\sum_\mathbf{X} 
	\frac{\langle \psi \vert \mathbf{X} \rangle \langle \mathbf{X} \vert S_i^+ S_j^-  \vert \psi \rangle}
	{\langle \psi \vert \psi \rangle} \nonumber \\
	& + & \frac{1}{4} (N_{\uparrow}-N_{\downarrow})^2 + \frac{1}{2} (N_{\uparrow} + N_{\downarrow})  \nonumber \\
	& = & \sum_{i=1}^{N_{\uparrow}} \sum_{j=N_{\uparrow}+1}^N 
	\sum_\mathbf{X} 
	p(\mathbf{X})
	{\langle \mathbf{X} \vert S_i^+ S_j^-  \vert \psi \rangle \over  \langle \mathbf{X} \vert \psi \rangle} \nonumber \\
	& + & \frac{1}{4} (N_{\uparrow}-N_{\downarrow})^2 + \frac{1}{2}N,
	\label{spinsquare}
\end{eqnarray}
where the operator $\vec{S}_i$ in the above equation acts on the spin component corresponding to  the electron position ${\mathbf r}_i$ of the 
configuration ${\mathbf X}$.
We can notice that 
\begin{equation}
p(\mathbf{X}) = \frac{\vert \langle \psi \vert \mathbf{X} \rangle \vert^2}{\sum_\mathbf{X} \vert \langle \psi \vert \mathbf{X} \rangle \vert^2},
\end{equation}
and that, by using QMC sampling, we generate configurations according to the probability density $p(\mathbf{X})$. Thus we can evaluate the above multidimensional integral by  directly sampling the estimator $S^2(\mathbf{X})$ that multiplies $p(\mathbf{X})$ in Eq.~(\ref{spinsquare}), namely
\begin{eqnarray}
S^2( \mathbf{X}) 	& = & \sum_{i=1}^{N_{\uparrow}} \sum_{j=N_{\uparrow}+1}^N 
  {	\langle \mathbf{X} \vert S_i^+ S_j^-  \vert \psi \rangle \over \langle \mathbf{X} \vert \psi \rangle}
    \nonumber \\
	& + & \frac{1}{4} (N_{\uparrow}-N_{\downarrow})^2 + \frac{1}{2}N.
	\label{spinsquared}
\end{eqnarray}

The content of the former equation can be evaluated efficiently as we will explain in the following. Indeed, the application of the operator $ S_i^+ S_j^-$ to the configuration $\mathbf{X}$ generates only a configuration 
$\mathbf{X}_{ij} = \{ ({\bm r}_1 \uparrow), \cdots ({\bm r}_i \downarrow), \cdots, ({\bm r}_j \uparrow), \cdots ({\bm r}_N \downarrow) \}$.
Considering $\mathbf{X}$ our sampled configuration and using the previously given definition of $\mathbf{X}_{ij}$,
we can recast Eq.~(\ref{spinsquared}) as
\begin{equation}
	 S^2 (\mathbf{X}) = 
	 \frac{1}{4} (N_{\uparrow}-N_{\downarrow})^2 + \frac{1}{2} N +
	\sum_{i=1}^{N_{\uparrow}} \sum_{j=N_{\uparrow}+1}^N 
	\frac{\langle \mathbf{X}_{ij} \vert \psi \rangle}{\langle \mathbf{X} \vert \psi \rangle}. 
	\label{eq:stot:formula}
\end{equation}
The only hard challenge of Eq.~(\ref{eq:stot:formula}) is the calculation of the  $N_{\uparrow} \times N_{\downarrow}$ ratios
\begin{equation}
r_{ij} = \frac{\langle \mathbf{X}_{ij} \vert \psi \rangle}{\langle \mathbf{X} \vert \psi \rangle} 
\end{equation}
for $i=1,2,\cdots, N_{\uparrow}$ and $j = N_{\uparrow}+1, N_{\uparrow}+2, \cdots, N$, that in our case this reads 
\begin{equation}
r_{ij} = \frac{J(\mathbf{X}_{ij})  \Psi_{AGP}(\mathbf{X}_{ij})}{J(\mathbf{X})  \Psi_{AGP}(\mathbf{X})}= r^{AGP}_{ij}r^{JF}_{ij}.
\label{ratios}
\end{equation}
The configurations $\mathbf{X}$ and $\mathbf{X}_{ij}$ differs for a spin flip of the electrons $i$ and $j$, but we can also consider $\mathbf{X}_{ij}$ as the configuration in which the electron $i$ evolved  to the position previously occupied by $j$ and viceversa. We can then calculate the ratios in Eq.~(\ref{ratios}) using a fast algebra to update two positions for the AGP and for the JF with a direct evaluation based on the Sherman-Morrison algebra and some simple manipulations, as discussed in details later on. 

It is also possible to calculate the value $S^2(\Lambda)$ of the $S^2$ operator in a sub-region of the space $\Lambda$.
For this quantity, we can obtain the similar expression to Eq.~(\ref{eq:stot:formula}):
\begin{equation}
	 S^2_{\Lambda} (\mathbf{X}) = 
	 \frac{1}{4} (N_{\uparrow}^{\Lambda}-N_{\downarrow}^{\Lambda})^2 + \frac{1}{2} N^{\Lambda} +
	\sum_{i = \{ \Lambda, \uparrow \}}
        \sum_{j = \{ \Lambda, \downarrow \}}
	\frac{\langle \mathbf{X}_{ij} \vert \psi \rangle}{\langle \mathbf{X} \vert \psi \rangle},
	\label{eq:stot:region}
\end{equation}
where $N^{\Lambda}_{\sigma}$ ($\sigma = \uparrow, \downarrow$) is the number of $\sigma$-electrons in the region $\Lambda$, $N^{\Lambda} = N_{\uparrow}^{\Lambda} + N_{\downarrow}^{\Lambda}$. The summation symbol over $i \in \{ \Lambda, \sigma\}$ indicates the sum for all $\sigma$-electron whose coordinate is in the region $\Lambda$.
Therefore, we can use same method described below. 

\subsubsection {The AGP contribution}

To calculate the AGP contribution to $r_{ij}$ we were able to find a slim and fast algebra making an extensive use of the Pfaffian properties \cite{Caracciolo2013474}. It was fundamental to find an efficient algebra to calculate the whole matrix of the ratios $r$ with a computational cost that is $O(N^3)$, by using mostly BLAS3 operations, thus avoiding that this computation could become the bottleneck of the whole procedure. In this way we could ensure the evaluation cost of $S^2$ to be comparable with the one of a typical  QMC cycle over all the $N$  electrons that is at  most  $O(N^3)$. Before describing the fast updating rules for the position of two electrons with a single move, we need to introduce some quantities fundamental for the calculation. 

Let us denote as $W^{-1}$ the inverse of $W$. This inverse  $W^{-1}$ can be computed 
from scratch for each configuration used to sample the spin square.
The electron  coordinates $\mathbf{r}_i$ are given  for $i=1,\cdots N$, but since the 
corresponding spin can  change  with respect to the  original choice 
($\uparrow$ for $i\le N_\uparrow$, and $\downarrow$ for $i>N_\uparrow$) due to the spin flips mentioned in the previous subsection,  we will consider explicitly 
the values of the spin here.

We then  define the matrix $\theta$ as
\begin{equation}
\theta_{ij}= f(\mathbf{r}_i \uparrow, \mathbf{r}_j \downarrow) + f(\mathbf{r}_i \downarrow, \mathbf{r}_j \uparrow)- f(\mathbf{r}_i \uparrow, \mathbf{r}_j \uparrow) -f(\mathbf{r}_i \downarrow, \mathbf{r}_j \downarrow).
\end{equation}
For the spin $\uparrow$ electrons we can define the vectors
\begin{equation}
	v_k^\uparrow = \left (
	\begin{matrix}
		f(\mathbf{r}_1 \uparrow, \mathbf{r}_k \uparrow) - f(\mathbf{r}_1 \uparrow, \mathbf{r}_k \downarrow)  \\
		f(\mathbf{r}_2 \uparrow, \mathbf{r}_k \uparrow) - f(\mathbf{r}_2 \uparrow, \mathbf{r}_k \downarrow)  \\
		\vdots \\
		f(\mathbf{r}_N \downarrow, \mathbf{r}_k \uparrow) - f(\mathbf{r}_N \downarrow, \mathbf{r}_k \downarrow)  \\
	\end{matrix}
	\right ),
\end{equation}
while for the spin $\downarrow$ we have instead
\begin{equation}
	v_l^\downarrow = \left (
	\begin{matrix}
		f(\mathbf{r}_1 \uparrow, \mathbf{r}_l \downarrow) - f(\mathbf{r}_1 \uparrow, \mathbf{r}_l \uparrow)  \\
		f(\mathbf{r}_2 \uparrow, \mathbf{r}_l \downarrow) - f(\mathbf{r}_2 \uparrow, \mathbf{r}_l \uparrow)  \\
		\vdots \\
		f(\mathbf{r}_N \downarrow, \mathbf{r}_l \downarrow) - f(\mathbf{r}_N \downarrow, \mathbf{r}_l \uparrow)  \\
	\end{matrix}
	\right ).
\end{equation}
We can use these vectors to build the $N\times N$ matrix
\begin{equation}
	V=\left( v_1^\uparrow v_2^\uparrow \cdots v_{N_\uparrow}^\uparrow  v_{N_\uparrow+1}^\downarrow  \cdots v_{N}^\downarrow  \right)= \left( V^\uparrow V^\downarrow \right),
\end{equation}
that allows us to define
\begin{equation}
	U=\left( U^\uparrow U^\downarrow \right)=W^{-1}V=\left( W^{-1}V^\uparrow W^{-1}V^\downarrow \right),
\end{equation}
and finally
\begin{equation}
	D=(V^\uparrow)^T U^\downarrow.
\end{equation}
Now we have all the ingredients that we need for our fast updating algebra, and upon 
application of Sherman-Morrison algebra,  we arrive  at the ratio 
\begin{eqnarray}
	r^{AGP}_{ij}&=& {\rm Pf}[W(\mathbf{X}_{ij}) ] /{\rm  Pf}[ W(\mathbf{X}) ] \nonumber \\
	&=&(1+U_{ii})(1+U_{jj})-U_{ij}U_{ji} - (\theta_{ij}+D_{ij})W^{-1}_{ij}.
\end{eqnarray}
We can notice that the preliminary calculation of the auxiliary matrices $\theta$, $V$, $U$ and $D$, 
including  the inversion of $W$, amounts to a total of $O(N^3)$ operations, while the calculation of the ratios is $O(N^2)$ once the matrices have been computed.

\subsubsection{The JF contribution}

In the JF that we introduced in the previous section only the two-body term of Eq.~(\ref{mb1}) has a spin dependence and thus only this part gives a contribution to the ratio. By simple substitution it is easy to prove that
\begin{eqnarray}
	r^{JF}_{ij} & = &\exp ( D_i -D_j +  u_{ee}(\mathbf{r}_i \uparrow, \mathbf{r}_j \downarrow) +  u_{ee}(\mathbf{r}_i \downarrow, \mathbf{r}_j \uparrow) \nonumber  \\
			& - & u_{ee}(\mathbf{r}_i \uparrow, \mathbf{r}_j \uparrow) - 
	u_{ee}(\mathbf{r}_i \downarrow, \mathbf{r}_j \downarrow) ),
\end{eqnarray}
where we have defined
\begin{equation}
	D_k=\sum_l u_{ee}(\mathbf{r}_l \sigma_l, \mathbf{r}_k \downarrow) - u_{ee}(\mathbf{r}_l \sigma_l, \mathbf{r}_k \uparrow).
\end{equation}
The whole operation has a $O(N^2)$ computational cost and so does not limit the calculation in terms of performances.

\section{Results and Discussion}

We apply this new approach for two types of systems: the first row high spin atoms (carbon, nitrogen and oxygen) and their diatomic molecules and the benzene molecule. The first ones still represent  useful benchmarks 
for the quantum chemistry approach and a reasonable description of their properties and binding energies requires very expensive multi-reference methods. It is therefore very interesting to test our approach to find if we are able to obtain a good description with a single Pfaffian {\it ansatz}. Benzene molecule on the other side is the most famous and important example of the RVB theory, so it represents a fundamental benchmark test for a method inspired by this theory.  In the following we will compare our results with exact available solutions, JSD WFs and with the JFVCAS multi determinant expansions for QMC. We will also show that our WF satisfies the size consistency both at VMC and DMC levels, a primary requirement if we want to use this approach for more challenging chemical studies.

\begin{figure}
        \centering
        \includegraphics[scale=0.85]{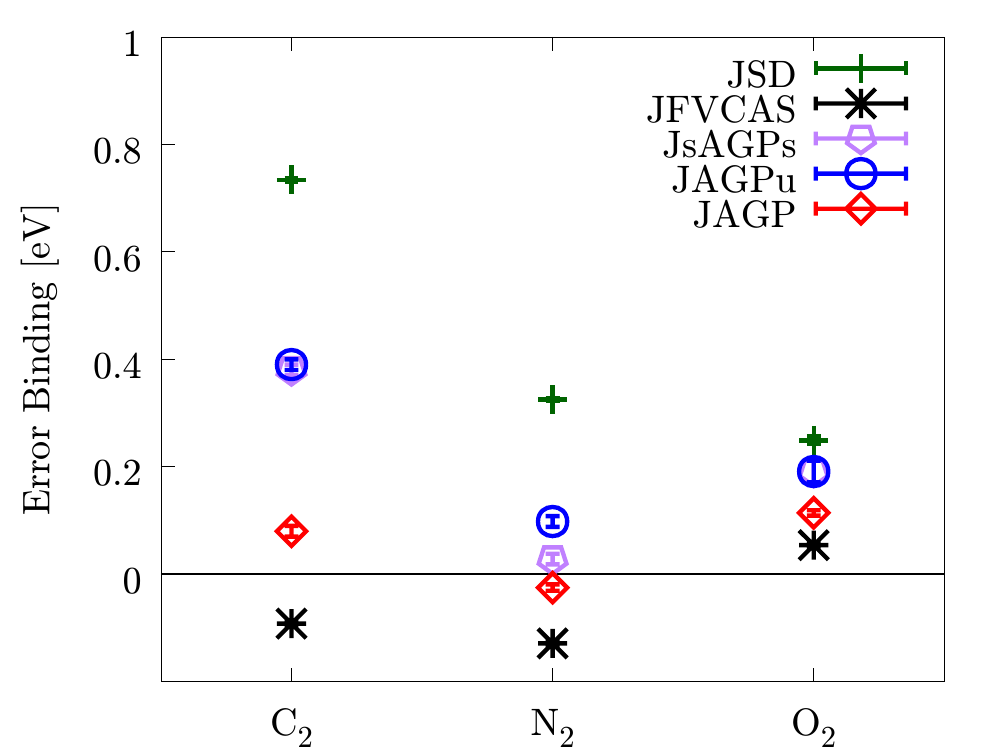}
        \caption{Comparison of the different DMC energies for different WFs. The results are shown for the three dimers described in this paper. The JFVCAS and the JSD results are taken from literature\cite{doi:10.1063/1.2908237}. }
        \label{comparison}
\end{figure}

\begin{table}
  \caption{Spin measures with different WFs for the  Carbon atom and dimer at VMC level.}\label{tab:cspins}
  \begin{tabular}{l c c c}
    \hline
    \hline
    \rule[-0.4mm]{0mm}{0.4cm}
    &  \multicolumn{2}{c}{$S^2$} & $2 \mu_B$ \\
    \rule[-0.4mm]{0mm}{0.4cm}
    &  Atom & Molecule & Moment $\parallel z$\\
    \hline
    \rule[-0.4mm]{0mm}{0.4cm}
    JsAGPs & 2.00 & 0.00  & 0.0005(4)\\
    \rule[-0.4mm]{0mm}{0.4cm}
    JAGPu  & 2.00534(3) &  0.1743(5)   & 0.5833(4)  \\
    \rule[-0.4mm]{0mm}{0.4cm}
    JsAGP  & 2.00418(5) &  0.2880(4)    & 0.7194(4)\\
    \rule[-0.4mm]{0mm}{0.4cm}
    JAGP  & 2.00542(1) &  0.0327(1)  & 0.0013(5) \\
    \rule[-0.4mm]{0mm}{0.4cm}
    Exact  & 2.00 & 0.00 & -  \\
    \hline
    \hline
\end{tabular}                   
\end{table}

\subsection{Carbon}

Carbon dimer is probably the most interesting example discussed in this study.  A full understanding of the behaviour of the carbon carbon interaction is still missing and the bond order of this molecule is still under debate\cite{Shaik2012QuadrupleBI}. 
The role of the spin fluctuations in this molecule has already been discussed\cite{genovese2019nature}, but we believe it is very instructive and so far represents the most important achievement of the JAGP WF. Indeed it is only thanks to the spin fluctuations that we can have a correct description of its dimer bond.

The carbon atoms have  spin triplet electronic configurations, and their mutual 
interaction leads to 
a singlet molecule.  As we can see from Fig. \ref{comparison} and table \ref{tab:carbon}, the JAGP not only improves the results of the JSD WF, but remarkably also the description given by the JsAGPs and JAGPu. The huge difference between the multideterminant expansion JFVCAS and the JSD binding energies helps to quantify the effect of the multi determinantal nature of this molecule, and this makes even more surprising the quality of the results obtained with a single JAGP WF that,  with a computational cost comparable to a SD, is already very close to the exact value.

\begin{table}
    \caption{Carbon Energies. The JsAGPs, JAGPu and JAGP results are calculated with an optimized ccpVTZ basis set. }\label{tab:carbon}   
    \begin{tabular}{l l l l}
      \hline
      \hline
      \multicolumn{4}{c}{\textbf{Carbon}}  \\
      \hline
      \rule[-0.4mm]{0mm}{0.4cm}
      & \multicolumn{1}{c}{Atom} & \multicolumn{1}{c}{Molecule} & \multicolumn{1}{c}{Binding}\\
      \hline
      \rule[-0.4mm]{0mm}{0.4cm}
      Source & \multicolumn{1}{c}{Energy$[H]$} & \multicolumn{1}{c}{Energy$[H]$} & \multicolumn{1}{c}{Energy$[eV]$} \\
      \hline
      \rule[-0.4mm]{0mm}{0.4cm}
      JSD  & -37.81705(6)${}^a$&  -75.8088(5)${}^a$   & 4.75(1) ${}^a$\\
      \rule[-0.4mm]{0mm}{0.4cm}
      JFVCAS  & -37.82607(5)${}^a$ &  -75.8862(2)${}^a$   & 6.369(6)${}^a$ \\
      \rule[-0.4mm]{0mm}{0.4cm}
      JsAGPs  & -37.8243(1)& -75.8611(2)& 5.78(1) \\
      \rule[-0.4mm]{0mm}{0.4cm}
      JAGPu  & -37.8263(1)& -75.8706(2)& 5.93(1)\\
      \rule[-0.4mm]{0mm}{0.4cm}
      JAGP  & -37.827965(3) & -75.88650(4)& 6.274(3)\\
      \rule[-0.4mm]{0mm}{0.4cm}
      JSD (DMC)  &-37.82966(4)${}^a$    & -75.8672(1)${}^a$   & 5.656(3)${}^a$ \\
      \rule[-0.4mm]{0mm}{0.4cm}
      JFVCAS (DMC) & -37.83620(1)${}^a$   & -75.9106(1)${}^a$  & 6.482(3)${}^a$ \\
      \rule[-0.4mm]{0mm}{0.4cm}
      JsAGPs (DMC)& -37.8364(1) &-75.8938(2)& 6.01(1) \\
      \rule[-0.4mm]{0mm}{0.4cm}
      JAGPu  (DMC)& -37.8364(1)& -75.8935(2)& 6.00(1)\\
      \rule[-0.4mm]{0mm}{0.4cm}
      JAGP (DMC) & -37.8363(1) & -75.9045(2) & 6.31(1) \\
      \rule[-0.4mm]{0mm}{0.4cm}
      Estimated Exact & -37.8450${}^b$  & -75.9265${}^c$  & 6.44(2)${}^{c,d}$ \\
      \hline
      \hline
     \multicolumn{4}{l}{${}^a$ Reference \cite{doi:10.1063/1.2908237}.}\\
     \multicolumn{4}{l}{${}^b$ Reference \cite{PhysRevA.47.3649}.}\\
     \multicolumn{4}{l}{${}^c$ Reference \cite{doi:10.1063/1.1869493}.}\\ 
     \multicolumn{4}{l}{${}^d$ A more recent estimate yields 6.39$eV$ }\\
      \multicolumn{4}{l}{(Cyrus Umrigar, private communication).}\\

    \end{tabular}			
  \end{table}

\begin{figure}
        \centering
        \includegraphics[scale=1]{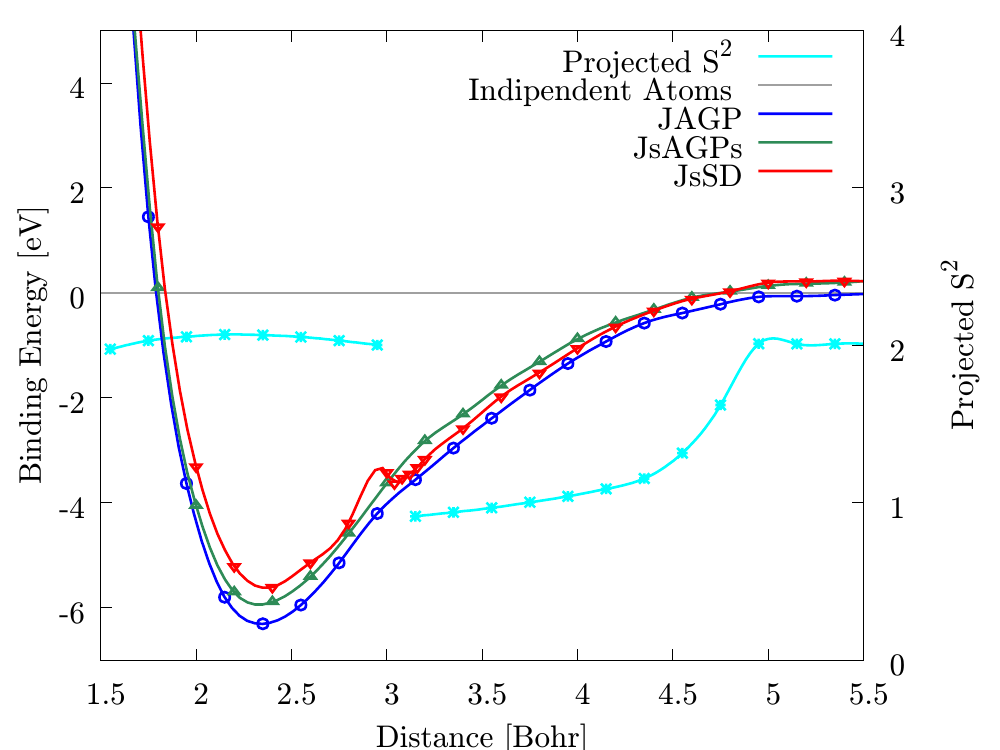}
        \caption{DMC energy dispersion of the carbon dimer: only the JAGP allows the system to be size consistent at large distance, namely it is able  to recover the energy and the expectation value of the $S^2$ operator of two isolated atoms. At bond distance however the carbon atoms maintain a large value of $S^2$. The sharp change of the projected $S^2$ value at around $3\ a.u.$ is probably  due to an avoided crossing of two energy levels belonging to 
	the same irreducible  representation, in agreement with  DMRG\cite{doi:10.1063/1.4905237}. Within LSDA this effect is reproduced by  a discontinuous change  in  the occupation of  the $\pi$ orbitals in the corresponding Slater determinant. Lines are guides to the eye.}\label{C2sizeconsistency}

\end{figure}

As already mentioned before, the explanation for the impressive improvement of the binding energy from JsAGPs and JAGPu to JAGP resides on the description of the strong spin fluctuations  in this molecule. The JAGP gives a very accurate picture of its magnetic properties as we can see from table \ref{tab:cspins}, giving results very close to $S^2=2$ for the atom and $S^2=0$ for the molecule. Conversely, by using the JsAGP (the AGP without spin dependent JF) and the JAGPu, we cannot recover the singlet from the broken symmetry initialization. Interestingly, as expected, the molecule does not have any magnetic moment on the $z$ direction, because it is an almost  perfect singlet. The atomic spins,
 localized around each atom,  point in opposite directions in order to form the singlet molecular state. Since there is no magnetic moment along $z$ we can measure its magnetic moment only by separately evaluating the $S^2$ in the two semi-infinite regions, each one containing a single atom, separated by a plane perpendicular to the molecular axis and at the same distance from the two atoms. In Fig. \ref{C2sizeconsistency} we show that even at bond distance there is a very strong magnetic moment around the atoms and, in this way, we can explain the strong effect of the zero point energy of the spin fluctuations described by the JAGP.

Moreover Fig. \ref{C2sizeconsistency} shows that only with the JAGP WF we have a size consistent solution with the molecule that recovers the energy of two independent atoms at large distance. This feature is fundamental if we want to use this WF to describe chemical reactions and perform  large scale simulations, with a size consistent behaviour at large distances.  
The importance of the  variational optimization of the wave function 
is particularly  evident in this small molecule. 
With the standard approach, by applying DMC to a SD 
taken by DFT (here obtained with Purdue and Zunger  LDA\cite{perdew1981self}), an unphysical 
level crossing in the occupation of  the $\pi$  molecular orbitals occurs
at around  $3\ Bohr$ distance, above which the $\pi$ bonding orbitals are only partially occupied. This  implies 
clear artifacts in the DMC energies.
We have verified that this level crossing is reproduced  with a standard DFT-LDA calculation by Gaussian16 A.03 revision~{\cite{gaussian16}} and an almost converged basis  set (the standard cc-pVQZ).
The level crossing has also been observed in Ref.{~\citenum{Sandeep_C2}}.
In our variational optimization instead, we have verified that 
it is important to start at large distance with the WF predicted by LSDA, otherwise a sizably higher energy is obtained. This effect is reflected also by the sharp change of the  projected $S^2$ at around $3\ Bohr$ distance (see Fig. \ref{C2sizeconsistency}), that could be  compatible with an avoided crossing between two energy  levels belonging to the same $^1\Sigma^+_g$ representation\cite{doi:10.1063/1.4905237}.

\subsection{Nitrogen}

Nitrogen is in some sense similar to the carbon case: also its dimer is indeed a singlet formed by two large spin $3/2$ atoms. 

As we can notice from Fig. \ref{comparison} and table \ref{tab:nitrogen}, at DMC level the JsAGPu and JAGP are both exact within chemical accuracy. 
All our calculations compares with the exact result better than the JFVCAS solution. Surprisingly, at VMC level the binding energies calculated with JAGP, JsAGPs and JAGPu are also very good. 

We remark that a  very powerful method, as the recently proposed Fermi Net\cite{pfau2019abinitio} (a neural network based WF), cannot reach the same precision in the binding energy even if the total energies of the molecule and atom are  the best  available ones. This clearly shows that all our ansatzs  allow a remarkable cancellation of errors, when computing the total energy differences  between the molecule and the two independent atoms.

\begin{table} 
  \caption{Nitrogen Energies. The JsAGPs, JAGPu and JAGP results are calculated with an optimized ccpVTZ basis set.}\label{tab:nitrogen}
  \begin{tabular}{l l l l l }
    \hline
    \hline
    \multicolumn{4}{c}{\textbf{Nitrogen}}  \\
     \hline
      \rule[-0.4mm]{0mm}{0.4cm}
      & \multicolumn{1}{c}{Atom} & \multicolumn{1}{c}{Molecule} & \multicolumn{1}{c}{Binding}\\
      \hline
      \rule[-0.4mm]{0mm}{0.4cm}
      Source & \multicolumn{1}{c}{Energy$[H]$} & \multicolumn{1}{c}{Energy$[H]$} & \multicolumn{1}{c}{Energy$[eV]$} \\
    \hline
    \rule[-0.4mm]{0mm}{0.4cm}
    JSD  & -54.5628(1)${}^a$  & -109.4520(5)${}^a$ &  8.88(1)${}^a$ \\
    \rule[-0.4mm]{0mm}{0.4cm}
    JFVCAS  & \multicolumn{1}{c}{-} & -109.4851(3)${}^a$ &  9.78(1)${}^a$\\
    \rule[-0.4mm]{0mm}{0.4cm}
    JsAGPs  & -54.55794(6)& -109.4781(7)& 9.856(3)\\
    \rule[-0.4mm]{0mm}{0.4cm}
    JAGPu & -54.55998(5)&  -109.48155(7)& 9.840(3) \\
    \rule[-0.4mm]{0mm}{0.4cm}
    JAGP  & -54.56633(5)& -109.49226(7)& 9.785(3) \\
    \rule[-0.4mm]{0mm}{0.4cm}
    JSD (DMC)  & -54.57587(4)${}^a$ &  -109.5039(1)${}^a$  & 9.583(3)${}^a$ \\
    \rule[-0.4mm]{0mm}{0.4cm}
    JFVCAS (DMC) & \multicolumn{1}{c}{-} &  -109.5206(1)${}^a$ &  10.037(3)${}^a$ \\
    \rule[-0.4mm]{0mm}{0.4cm}
    JsAGPs (DMC)& -54.5765(1)& -109.5164(2)& 9.88(1)\\
    \rule[-0.4mm]{0mm}{0.4cm}
    JAGPu (DMC)& -54.5767(3)& -109.5140(2)& 9.81(1)\\
    \rule[-0.4mm]{0mm}{0.4cm}
    JAGP (DMC) & -54.57709(9)& -109.5192(1)& 9.933(6)\\
    \rule[-0.4mm]{0mm}{0.4cm}
    Fermi Net &    -54.58882(6)${}^b$ & -109.5388(1)${}^b$ & 9.828(5)${}^b$ \\
    \rule[-0.4mm]{0mm}{0.4cm}
    Estimated Exact & -54.5892${}^c$  &   -109.5427${}^d$ &  9.908(3)${}^d$ \\
    \hline
    \hline
     \multicolumn{4}{l}{${}^a$ Reference \cite{doi:10.1063/1.2908237}.}\\
     \multicolumn{4}{l}{${}^b$ Reference \cite{pfau2019abinitio}.}\\
     \multicolumn{4}{l}{${}^c$ Reference \cite{PhysRevA.47.3649}.}\\
     \multicolumn{4}{l}{${}^d$ Reference \cite{doi:10.1063/1.1869493}.}\\
\end{tabular}			
\end{table}

\begin{figure}
        \centering
        \includegraphics[scale=1]{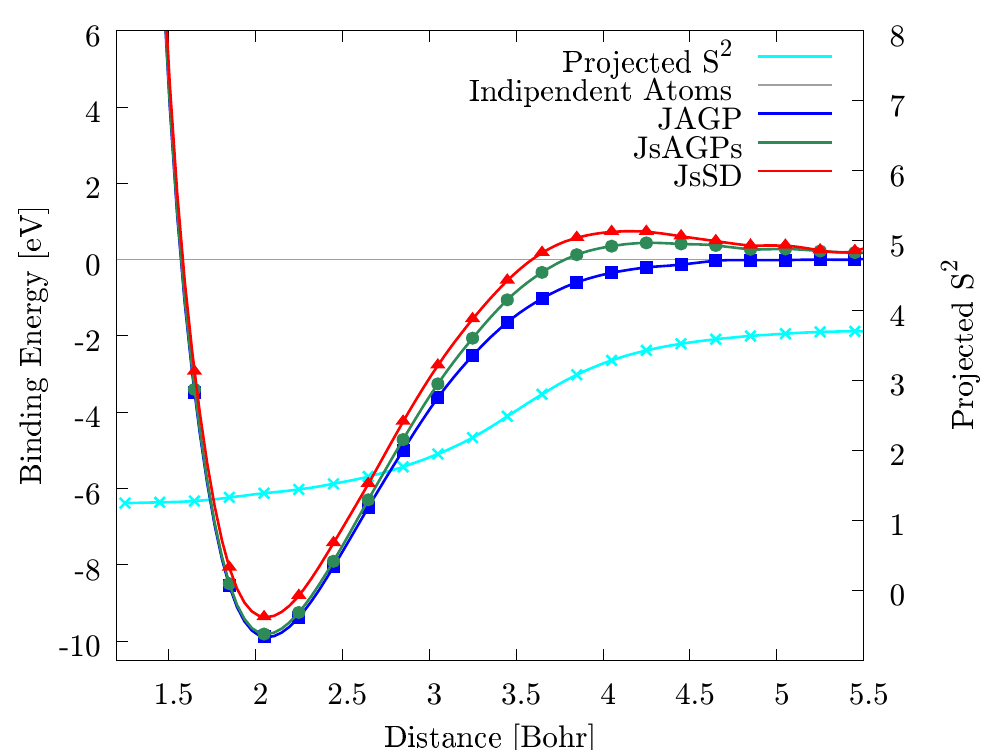}
        \caption{DMC energy dispersion of the nitrogen dimer: only the JAGP appears to be perfectly size consistent, thus   recovering  at large interatomic distance the energy and the expectation value of the $S^2$ operator
        of two isolated atoms. At bond distance however the nitrogen atoms have a smaller value of $S^2$, in contrast to what observed for the carbon dimer. Lines are guides to the eye.}
        \label{N2sizeconsistency}
\end{figure}

In this case, however, the difference between JAGP and JsAGPs$/$JAGPu is much smaller than in the previous case and should be related to a less important role of the spin fluctuations and also  to a smaller magnetic moment of the atoms at equilibrium distance. By repeating the reasoning done for the carbon dimer, we can quantify the magnetic moment from the $S^2$ value in the semi-infinite region separated by a plane perpendicular to the axis of the molecule and equidistant from the atoms. Looking at Fig. \ref{N2sizeconsistency} we can see that, at bond distance, the $S^2$ of the atom is much smaller than the one of an independent atom and therefore, even if the nitrogen atom has a large spin, when it is forming a dimer it does not give rise to a strong antiferromagnetism. 

Also in this case it is important to notice that the JAGP solution is size consistent both in energy and spin. Despite the very good description at bond distance provided by the JsAGPs, we notice from Fig.  \ref{N2sizeconsistency} that it is not perfectly size consistent.
Within our approach a fully consistent picture and a very accurate dispersion is possible only by means of the JAGP ansatz, that is able to work properly also in the strong correlation regime, namely at large interatomic distance.

\subsection{Oxygen}

The oxygen is very different from the previous cases but nevertheless very interesting for different reasons. The oxygen dimer consists of two triplet  atoms, but this time the molecule is a triplet. 
There are small atomic magnetic moments   in the GS of the oxygen molecule, but 
the role of 
the magnetic interaction remains important, as shown by the application of the JAGP ansatz. 
In this case it looks that   the interaction of parallel spins electrons is particularly important, 
and this can be   described by the JAGP ansatz  more accurately than  the  corresponding 
JsAGPs and JAGPu ones, as discussed in  the previous sections.  Thus  we expect to recover with  the JAGP some correlation that we miss when we simplify the ansatz by using the unpaired orbitals in the JsAGPs and in the JAGPu WFs.

\begin{table}
  \caption{Oxygen Energies. The JsAGPs, JAGPu and JAGP results calculated with an optimized ccpVTZ basis set.}\label{tab:oxygen}
  \begin{tabular}{l l l l }
    \hline
    \hline
    \multicolumn{4}{c}{\textbf{Oxygen}}  \\
    \hline
         \rule[-0.4mm]{0mm}{0.4cm}
      & \multicolumn{1}{c}{Atom} & \multicolumn{1}{c}{Molecule} & \multicolumn{1}{c}{Binding}\\
      \hline
      \rule[-0.4mm]{0mm}{0.4cm}
      Source & \multicolumn{1}{c}{Energy$[H]$} & \multicolumn{1}{c}{Energy$[H]$} & \multicolumn{1}{c}{Energy$[eV]$} \\
    \hline
     \rule[-0.4mm]{0mm}{0.4cm}
    JSD  & -75.0352(1)${}^a$  & -150.2248(5)${}^a$ & 4.20(1)${}^a$ \\
    \rule[-0.4mm]{0mm}{0.4cm}
    JFVCAS  & \multicolumn{1}{c}{-}  & -150.2436(2)${}^a$ & 4.713(8)${}^a$ \\
    \rule[-0.4mm]{0mm}{0.4cm}
    JsAGPs  & -75.0268(3)& -150.2372(6)& 5.00(3) \\
    \rule[-0.4mm]{0mm}{0.4cm}
    JAGPu &  -75.0339(3)& -150.2503(5)& 4.97(3)\\
    \rule[-0.4mm]{0mm}{0.4cm}
    JAGP  &  -75.0346(2)& -150.2572(4)& 5.11(2)\\
    \rule[-0.4mm]{0mm}{0.4cm}
    JSD (DMC)  & -75.05187(7)${}^a$  & -150.2872(2)${}^a$  &  4.992(7)${}^a$\\
    \rule[-0.4mm]{0mm}{0.4cm}
    JFVCAS (DMC) &  \multicolumn{1}{c}{-}   & -150.29437(9)${}^a$  & 5.187(5)${}^a$ \\
    \rule[-0.4mm]{0mm}{0.4cm}
    JsAGPs (DMC)& -75.0518(3) & -150.2894(3) & 5.06(2)\\
    \rule[-0.4mm]{0mm}{0.4cm}
    JAGPu (DMC)&  -75.0519(3) &  -150.2902(4) & 5.06(2)\\
    \rule[-0.4mm]{0mm}{0.4cm}
    JAGP (DMC) & -75.05289(7) & -150.2942(1) & 5.127(5)\\
    \rule[-0.4mm]{0mm}{0.4cm}
    Estimated Exact & -75.0673${}^b$ & -150.3724${}^c$   & 5.241${}^c$ \\
    \hline
    \hline
     \multicolumn{4}{l}{${}^a$ Reference \cite{doi:10.1063/1.2908237}.}\\
     \multicolumn{4}{l}{${}^b$ Reference \cite{PhysRevA.47.3649}.}\\
     \multicolumn{4}{l}{${}^c$ Reference \cite{doi:10.1063/1.1869493}.}\\
\end{tabular}			
\end{table}%

\begin{figure}
        \centering
       \includegraphics[scale=1]{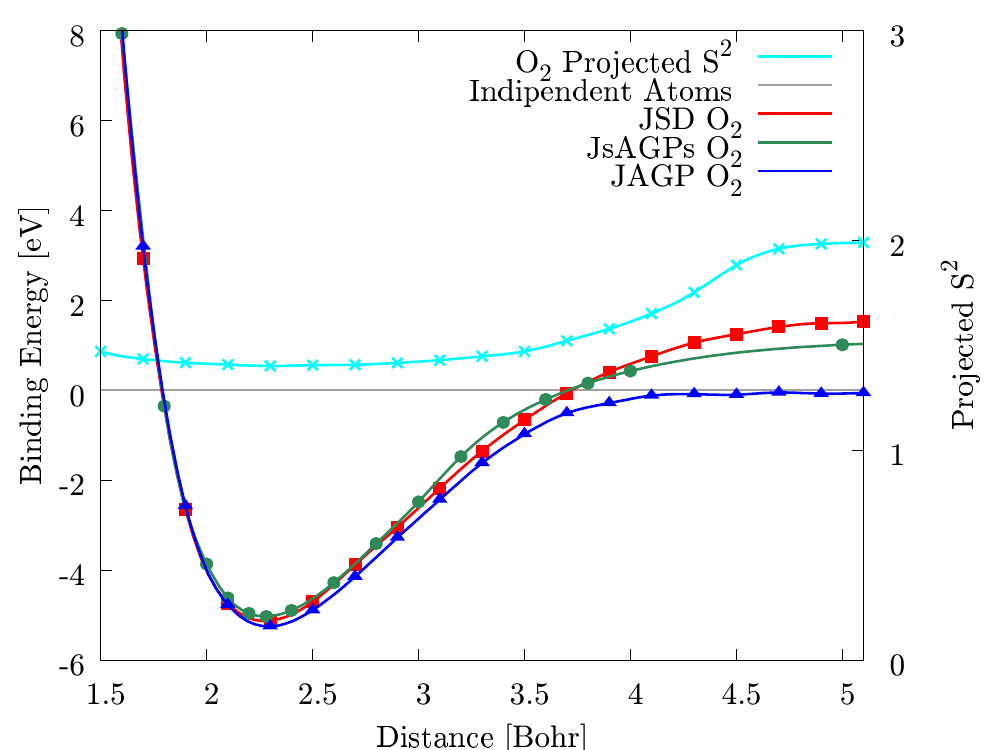}
        \caption{DMC energy dispersion of the oxygen dimer with the JAGP, JsAGP \cite{doi:10.1063/1.4885144} and JSD (with the SD obtained from DFT calculations): at large distance only the JAGP WF is size consistent. In the plot also the expectation value of the projected $S^2$ operator on the atoms for the JAGP that recovers at large distance the value of two isolated atoms. Lines are guides to the eye.}
        \label{O2sizeconsistency}
\end{figure}

By looking at Fig. \ref{comparison} and table \ref{tab:oxygen} we can see that, at DMC level, the energies obtained with the JAGP WF are extremely good even  for the oxygen dimer.
In this case the correct description of the triplet pairing correlations, possible within the JAGP {\it ansatz}, appears to be fundamental. Indeed the final result is so accurate that the binding energy is comparable to the one obtained with the multi-determinant JFVCAS WF. It is even more surprising that the absolute energies of the atom and molecule are very close where not even better than the ones provided by the multi-determinat expansion both at VMC and DMC level. We have to point out, however, that within JFVCAS method it is not possible to improve the JSD atom \cite{doi:10.1063/1.2908237} and that the binding energy slightly  better than the JAGP one  derives  from  the poorer quality of the atom rather than a better description of the molecule. 

The problem of the size consistency for the oxygen dimer is absolutely non trivial and even more complicated than the previous cases. Starting from bond distance we have a molecule of spin one and fixed projection $S_z=1$ but we have to recover the behaviour of two independent atoms.  This means that, by keeping the projection $S_z=1$ constant, while separating the atoms far apart, we have to recover the correct atoms of spin one and thus we need to have one atom with the spin oriented in a direction perpendicular to the $z-$axis. This is impossible for the JsAGPs and the JAGPu but allowed by the JAGP, a remarkable and absolutely non trivial feature of  this WF. As we can see from Fig. \ref{O2sizeconsistency}, at large distance only with the JAGP the system recovers the energy and the spins of the independent atoms, showing that, by means of  our advanced optimization tools, it is possible to dramatically change the WF  up to the point of rotating completely the spin of an atom.

\subsection{Benzene}

The benzene molecule represents one of the most successful example of the RVB theory with the carbon-carbon bonds resonating among  several valence bond configurations, e.g. Kekul\'e and Dewar. QMC methods are able to provide a very good description of this important molecule \cite{doi:10.1063/1.4930137, doi:10.1063/1.2746035}, and thus it is interesting to check whether, with our new approach, we can obtain a very accurate result. In particular in table~\ref{tab:benzene} we compare the results obtained by JSD, JAGP, JAGPu and JsAGPs  WF, showing that all  the  results obtained with a  pairing  function (from JsAGPs to JAGP) 
provide a very good estimate of the absolute energies, noticeably  improving the results of the JSD. Moreover the corresponding atomization energies are extremely accurate at the DMC level,  whereas the JSD largely overestimate it. It is finally interesting to notice that, even if there is a sizeable gain in terms of absolute energy with  our best ansatz, i.e. the  JAGP,
it is not clear why  this systematic improvement does not sizeably affect  the atomization energy,   likewise this could be almost  converged to the exact value. This 
might  be in principle explained because, at present,  the accuracy of the state of the art ''estimated exact''  calculation is probably not enough  to establish an energy difference  $<<0.1eV$. For instance the zero point energy (ZPE) has been  estimated by DFT\cite{doi:10.1002/bbpc.19900940121} and some work is certainly  necessary to clarify this issue, e.g. by calculating the ZPE directly with QMC.

We remark  here   that  the JsAGPS description of the  benzene molecule  is already very accurate and it is not improved by the JAGP. This is probably due  to the lack of any sizeable spin moment around any atom composing this molecule. Indeed the $S^2$ value calculated for the JAGP and JAGPu solutions are $0.032(1)$ and $0.0123(7)$,  respectively, proving that any local  magnetic moment  is almost completely melted during the optimization, despite its non zero initialization. We conclude therefore that in the benzene molecule the spin fluctuations are not  relevant and the use of the Pfaffian leads only  to a marginal improvement of the total energy while the molecule is correctly 
described by  a  perfect singlet RVB ansatz given by the JsAGPs, in agreement with the classical RVB picture  by L. Pauling\cite{bk:pauling}.
\begin{table} 
  \caption{Benzene Energies}\label{tab:benzene}
  \begin{tabular}{l l l c  }
    \hline
    \hline
    \multicolumn{4}{c}{\textbf{Benzene}}  \\
     \hline
      \rule[-0.4mm]{0mm}{0.4cm}
      & \multicolumn{1}{c}{C atom${}^a$} & \multicolumn{1}{c}{Molecule} & \multicolumn{1}{c}{Atomization Energy}\\
      \hline
      \rule[-0.4mm]{0mm}{0.4cm}
      Source & \multicolumn{1}{c}{Energy$[H]$} & \multicolumn{1}{c}{Energy$[H]$} & \multicolumn{1}{c}{Energy$[eV]$} \\
      \hline
            \rule[-0.4mm]{0mm}{0.4cm}
	    JSD	& -37.8074(1) & -232.0261(3) &  59.37(1) \\
	    \rule[-0.4mm]{0mm}{0.4cm}
	    JsAGPs & -37.82383(4) & -232.0805(3) & 58.166(8) \\
             \rule[-0.4mm]{0mm}{0.4cm}
	    JAGPu  & -37.82651(5) & -232.0900(3) & 57.986(8) \\
             \rule[-0.4mm]{0mm}{0.4cm}
 	     JAGP & -37.82921(4) & -232.1060(2) & 57.982(7) \\
	  	    \rule[-0.4mm]{0mm}{0.4cm}
	    JSD(DMC) & -37.8299(1) & -232.1879(6) & 60.09(2) \\
	      \rule[-0.4mm]{0mm}{0.4cm}
	    JsAGPs(DMC) & -37.8368(1) & -232.1947(6) & 59.16(2) \\
	      \rule[-0.4mm]{0mm}{0.4cm}
	    JAGPu(DMC) & -37.8367(1) & -232.1943(6) & 59.16(2) \\
	     \rule[-0.4mm]{0mm}{0.4cm}
	    JAGP(DMC) & -37.83751(9) & -232.1998(5) & 59.18(2) \\
	    \rule[-0.4mm]{0mm}{0.4cm}
	    Estimated Exact & -37.8450${}^b$ & -232.250(1) & 59.32(2)${}^c$ \\
    \hline
    \hline
    \multicolumn{4}{l}{${}^a$ Calculated with the same basis set used for the benzene molecule.} \\ 
    \multicolumn{4}{l}{${}^b$ Reference \cite{PhysRevA.47.3649}.}\\
    \multicolumn{4}{l}{${}^c$ Reference \cite{doi:10.1002/bbpc.19900940121}.} \\ 

  \end{tabular}			
\end{table}

\section*{Conclusions}

In this work we have proposed a new WF for QMC calculations given by the most general fermionic pairing function {\it ansatzs }in combination with a spin JF that provides a very rich description of the electronic correlation by  means of  a  bosonic pairing function complementary to the fermionic one. With a computational cost comparable to a SD we were able to improve not only the results achieved with a simple JSD but also with JsAGPs and JAGPu, reaching a level of accuracy comparable to the one obtained with the multi-determinant JFVCAS WF. The powerful optimization techniques are probably the keys to explain the remarkable improvement we obtained with this WF, compared to previous attempts\cite{PhysRevLett.96.130201,PhysRevB.77.115112}. In particular we have shown that the JAGP {\it ansatz }provides a very accurate description of high spin atoms and their dimers and that it is size consistent. This should increase the number of possible applications, providing  a  reasonably accurate and computationally feasible tool for studying chemical reactions. 
The triplet correlations have proven to be necessary to take into account correctly the zero point energy of the spin fluctuations that we can now correctly describe thanks to a physical and accurate setup obtained by orienting the atomic magnetic moments of the AGP in the direction perpendicular to the  spin quantization axis chosen for the JF. For this reason we have obtained a very good description of the carbon and nitrogen dimers, remarkably even when the first molecule was found to be very poorly described by  the JsAGPs and the JAGPu. Moreover it is only thanks to the presence of the triplet correlations that we were able to improve the description of the oxygen dimer. a strongly correlated triplet molecule  with an highly entangled spin interaction among  the atoms.   
Comparison with other methods different from QMC is shown in App.~\ref{appDMRG}.
Our  QMC variational energy is much  better than state of the 
art quantum chemistry methods that seem to be affected by strong basis set errors  even  when considering only energy differences. 
For instance the total energy 
difference $\Delta E$ 
at $R=4.2a.u.$ and $R=2.11a.u.$ in Tab.~\ref{nitrogen_disp} 
should be close to the estimated exact binding  energy (i.e. $\simeq 9.91ev$ from Ref.{\citenum{doi:10.1063/1.1869493}}), at most weakly corrected by the residual dispersive interaction. Both  DMRG and MRCI clearly miss more than 1eV with the DZ basis, i.e. $\Delta E \simeq 8.49 eV$.
In order to show more clearly that the discrepancy between our DMC results 
and DMRG and MRCI is actually an artifact of the small basis, we have carried out UCCSD(T) calculation both  for small ($\Delta E = 8.6eV$)  and large 
($\Delta E=9.55 eV$) basis set, and, as expected,
our calculation ($\Delta E= 9.63$) is much more in agreement 
with the most accurate large basis set calculation.
In any event our binding energy for $N_2$  ($9.933\pm 0.006eV$)  
is surprisingly more accurate than the best state of the 
art calculation with CCSD(T) ($9.73eV$ from the Computational Chemistry Comparison and Benchmark DataBase~{\cite{NIST_data}}), implying that 
, most  likely, our results should be considered the state art for the 
full dispersion curve of these small molecules.

Finally we demonstrated that for the benzene dimer the JAGP is able to provide a very accurate atomization energy, though it is not clear in this case whether the triplet correlations are crucial for a highly accurate calculation. 
 However it is important to highlight 
that the accuracy in the binding energy is always much  better than the accuracy in the  total energy 
and that therefore there exists  always a remarkable cancellation of errors in the total energy differences. This feature indeed is fundamental for a compact {\it ansatz} like the JAGP and it challenges  other very expensive highly correlated methods, even when these are able to achieve almost exact total energies, as it was the case for  the Fermi Net approach  to the Nitrogen dimer.

The relatively low computational cost of QMC combined with  powerful optimization techniques, allowing 
a reasonably  large number of variational parameters, make this approach ideal for studying systems even much larger than the ones considered in this work. 
Indeed, we believe that the paradigm presented in this paper could represent in the future a very powerful tool to investigate the electronic structure of interesting chemical compounds and physical systems where the spin interaction may play an important role, that in turn may  be  a number 
much larger than previously believed, as we have  presented here the $C_2$ molecule, as the  very first and remarkable  example of an antiferromagnetic  chemical bond. 

\begin{acknowledgement}
We acknowledge PRIN-2017  for financial support  and CINECA PRACE-2019 for computational resources.
Parts of numerical simulations have been done on the HOKUSAI supercomputer at RIKEN (Project ID: 19011 and 19030).
K.N. is grateful for the facilities of Research Center for Advanced Computing Infrastructure at Japan Advanced Institute of Science and Technology (JAIST). K.N. also acknowledges a financial support from the Simons Foundation and that from Grant-in-Aid for Scientific Research on Innovative Areas (No.~16H06439).
\end{acknowledgement}

\bibliography{Bibliography}

\appendix

\section{Energy Dispersion Comparison}
\label{appDMRG}
\begin{figure}
        \centering
       \includegraphics[scale=1]{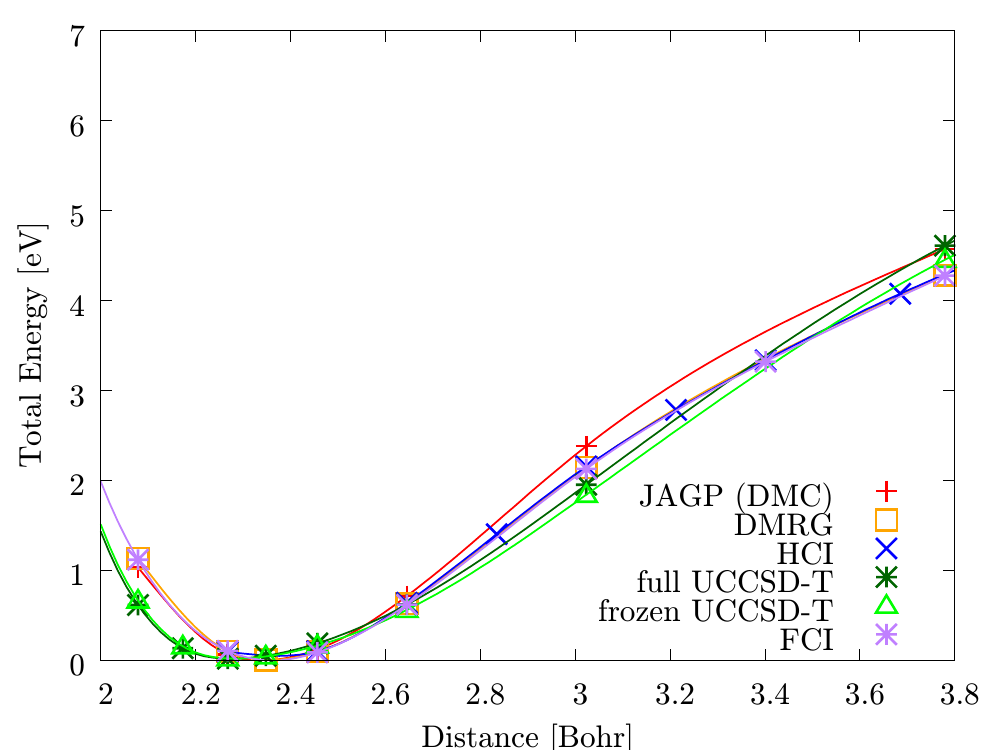}
        \caption{Energy dispersion of the carbon dimer calculated with JAGP (DMC), UCCSD-T (ccpV5Z), DMRG\cite{doi:10.1063/1.4905237}, HCI\cite{doi:10.1063/1.4998614} and FCI\cite{doi:10.1063/1.3624383}. Lines are guides to the eye.}
        \label{morse}
\end{figure}

Comparison between the energy dispersion calculated with DMC JAGP WF and unrestricted single reference coupled cluster (UCCSD-T) with ccpVDZ and ccpV5Z basis sets. We further compared the carbon energy dispersion with density matrix renormalization group (DMRG), heat-bath configuration interaction (HCI) and full configuration interaction (FCI) from literature\cite{doi:10.1063/1.4905237, doi:10.1063/1.4998614,doi:10.1063/1.3624383}, and the nitrogen dispersion with  multi-reference coupled cluster (MRCC) and DMRG\cite{doi:10.1063/1.1783212}.  U-CCSD(T) calculations were performed using Gaussian16 A.03 revision with the counterpoise correction, with the frozen-core approximation and the full-core correlation~{\cite{gaussian16}}. Table \ref{carbon_disp} and Fig.~\ref{morse} show that there 
are  significant discrepancies between different  methods  in the Carbon dimer dispersion curve at large distances.
However one has to consider that, even in a quadruple zeta basis the binding 
energy $D_e=6.22eV$\cite{doi:10.1063/1.3624383} is about 6mH lower than the estimated  exact  one and therefore if we reference all the curves at the bond length minimum energy, as reported in the mentioned figure,  a method that is supposed to be weakly dependent on the 
basis, as our DMC, should be slightly higher in energy at large  distance, provided it remains close to the  exact dispersion energy curve.
Moreover there may be sizeable corrections due to the frozen core approximation 
employed by DMRG, HCI and  FCI.  
We have indeed verified that they are non negligible  in the UCCSD-T calculation, implying that core-valence interaction can lead to a further non-parallelity error of about 3mH (see Fig.\ref{morse}).
Core-valence interaction   is   considered in our DMC  calculations simply because, within this technique,  it is not possible to employ the frozen core approximation. 
Nevertheless it is clear that our results may have some  error, but it  
is remarkable that  if we use 
the corresponding energy values for computing  the zero point energy (ZPE) of the dimer we find excellent
agreement with the 
experimental value, given by  $0.1146\ eV$\cite{doi:10.1063/1.2436891}. 
Indeed the ZPE calculated values, using  a standard fit with a quartic polynomial close to the equilibrium distance, are 
 $0.1153(6)\ eV$, $0.108\ eV$, $0.106\ eV$, $0.112\ eV$, $0.114\ eV$ and $1133(3)\ eV$  
 for our  DMC, UCCSD-T full core and frozen core, DMRG, HCI and FCI, respectively.
 In summary, by taking into account all possible sources of error, 
 we believe that our results are in reasonable agreement we the expected ''exact result'' converged  in the complete basis set limit and with full core-valence interaction taken into account. 
 Indeed we believe that 
 only a more direct comparison with experiments or 
 a full core  FCI/DMRG or HCI extrapolated to the complete basis set limit 
 can further improve the accuracy of the dispersion curve.

\begin{table} 
  \caption{Carbon Energy Dispersion [Hartree]. The JAGP results were obtained with the optimized ccpVDZ basis set (as explained in section \ref{sec:basis}), the DMRG results with the ccpVQZ basis, the HCI with ccpV5Z basis set, the FCI with ccpVQZ basis set,  whereas the UCCSD-T ones, both full and frozen core, are shown for ccpV5Z basis sets.}\label{carbon_disp}
  \begin{tabular}{l l l l l l l}
    \hline
    \hline
         \rule[-0.4mm]{0mm}{0.4cm}
      & \multicolumn{6}{c}{Numerical Technique} \\
      \hline
      \rule[-0.4mm]{0mm}{0.4cm}
      Distance &  \multicolumn{1}{c}{ JAGP (DMC)} &  \multicolumn{1}{c}{DMRG} & \multicolumn{1}{c}{HCI} & \multicolumn{1}{c}{UCCSD-T$_{frozen}$} & \multicolumn{1}{c}{UCCSD-T$_{full}$} &  \multicolumn{1}{c}{FCI}\\
      \hline
            \rule[-0.4mm]{0mm}{0.4cm}
	    2.0787 & -75.86652(3) & -75.76125$^b$ &-75.76701$^c$ & -75.76085& -75.78683 & -75.7624$^d$ \\
            \rule[-0.4mm]{0mm}{0.4cm}
	    2.2677 & -75.90207(3)& -75.79924$^b$ &-75.80461$^c$ & -75.78450& -75.80878 & -75.7987$^d$\\
            \rule[-0.4mm]{0mm}{0.4cm}
	    2.3480 & -75.90456(3)& -75.80269$^b$ &-75.80786$^{a,c}$ & -75.78370& -75.80754 &-75.8025$^{d}$\\
            \rule[-0.4mm]{0mm}{0.4cm}
	    2.4566 & -75.90008(3)& -75.79937$^b$ &-75.80444$^c$ & -75.77928& -75.80247 &-75.7993$^{d}$\\
            \rule[-0.4mm]{0mm}{0.4cm}
	    2.6456 & -75.87825(4)& -75.77937$^b$ &-75.78460$^c$ & -75.76465& -75.78664 &-75.7798$^{d}$\\
            \rule[-0.4mm]{0mm}{0.4cm}
	    3.0235 & -75.81700(8)& -75.72405$^b$ &-75.72895$^c$ & -75.71762 &-75.73765 & -75.7243$^{d}$\\
            \rule[-0.4mm]{0mm}{0.4cm}
	    3.7794 & -75.73649(8)& -75.64560$^b$ &-75.65043$^{a,c}$ & -75.62162 &-75.63996 &-75.6454$^{d}$\\
    \hline
    \hline
    \multicolumn{6}{l}{${}^a$ Interpolated.} \\ 
    \multicolumn{6}{l}{${}^b$ Reference \cite{doi:10.1063/1.4905237}.}\\
    \multicolumn{6}{l}{${}^c$ Reference \cite{doi:10.1063/1.4998614}.}\\
    \multicolumn{6}{l}{${}^d$ Reference \cite{doi:10.1063/1.3624383}.}\\
  \end{tabular}			
\end{table}

\begin{table} 
\caption{Nitrogen Energy Dispersion [Hartree]. The JAGP results were obtained with the optimized ccpVDZ basis set (as explained in section \ref{sec:basis}), the DMRG and MRCC results with the ccpVDZ basis, whereas the corresponding UCCSD-T ones are shown also for a much larger basis (ccpV5Z), resulting in much better agreement with the present DMC results.}\label{nitrogen_disp}
  \begin{tabular}{l l l l l l}
    \hline
    \hline
         \rule[-0.4mm]{0mm}{0.4cm}
  
      & \multicolumn{5}{c}{Numerical Technique} \\
      \hline
      \rule[-0.4mm]{0mm}{0.4cm}
      Distance &  \multicolumn{1}{c}{JAGP (DMC)} &  \multicolumn{1}{c}{DMRG} &  \multicolumn{1}{c}{MRCC} & \multicolumn{1}{c}{UCCSD-T (DZ)} & \multicolumn{1}{c}{UCCSD-T (5Z)}\\
      \hline
            \rule[-0.4mm]{0mm}{0.4cm}
	    2.118 & -109.51694(5) & -109.27833$^b$ & -109.27683$^b$  & -109.27652 & -109.41303$^a$\\
            \rule[-0.4mm]{0mm}{0.4cm}
	    2.4 & -109.46459(6) & -109.23838$^b$ & -109.23687$^b$ & -109.23202 & -109.35926\\
            \rule[-0.4mm]{0mm}{0.4cm}
	    2.7 & -109.37935(6) & -109.16029$^b$ & -109.15895$^b$ & -109.14731 &-109.26936\\
            \rule[-0.4mm]{0mm}{0.4cm}
	    3.0 & -109.29961(6) & -109.08619$^b$ & -109.08442$^b$ &-109.06570 &-109.18331\\
            \rule[-0.4mm]{0mm}{0.4cm}
	    3.6 &  -109.19745(6)& -108.99489$^b$ & -108.99272$^b$ &-108.97982 &-109.08833\\
            \rule[-0.4mm]{0mm}{0.4cm}
	    4.2 &  -109.16376(7)&  \multicolumn{1}{c}{-} & -108.96471$^b$ &-108.96002 &-109.06204\\
    \hline
    \hline
    \multicolumn{4}{l}{${}^a$ Interpolated.} \\ 
    \multicolumn{4}{l}{${}^b$ Reference \cite{doi:10.1063/1.1783212}.}\\
  \end{tabular}			
\end{table}

\section{Diagonalization of a skew symmetric generally  complex matrix $\lambda$}

In the following we will discuss a general 
procedure to transform 
a generic complex antisymmetric matrix 
into a canonical Youla's form that represents the equivalent  
of the standard diagonalization  of Hermitian matrices.
This is obtained by means of an appropriate  unitary matrix $U$  
defined by an orthonormal set of states that we will call in the following  MOs.

Given a  $\bar{N} \times \bar{N}$ antisymmetric matrix $\lambda$, our goal is to identify a set of $p$  paired states  $\{ (\phi_j^1,\phi_j^2)\}$ of orthonormal MOs, such that  $p\le \bar N$ is even and 
\begin{equation}
  \lambda \phi_j^1=a_j \phi_j^2
  \label{orth1}
\end{equation}
\begin{equation}
  \lambda \phi_j^2=-a_j \phi_j^1,
  \label{orth2}
\end{equation}
where the LHS of the  above  equations indicate standard 
matrix vector products, with shorthand notations adopted also in the remaining part of this appendix.
In this basis we can write any skew symmetric matrix
$\lambda$ in the canonical Youla's form:
\begin{equation}
  \lambda_{MO}= 
  \left (
  \begin{matrix} 
    0 & a_1 & 0 & 0 & \cdots & 0 \\ 
    -a_1 & 0 & 0 & 0 & \cdots & 0 \\
    0 & 0 & 0 & a_2  & \cdots & 0 \\
    0 & 0 & -a_2 & 0  & \cdots & 0 \\
    \vdots & \vdots & \vdots & \vdots & \ddots & \vdots 
  \end{matrix}
  \right ),
  \label{diaglambda}
\end{equation}
using only $p/2$ strictly positive  parameters $a_j$. These 
ones play the same role of the eigenvalues for an ordinary Hermitian matrix 
and henceforth we will use this name for them, even if the matrix $\lambda_{MO}$ is not diagonal 
but  represents the simplest non vanishing skew-symmetric matrix.

The transformation of the  original matrix $\lambda$ to  the corresponding canonical Youla's form 
by means of an appropriate unitary transformation $ \lambda = U^* \lambda_{MO} U^\dag$ 
provides us also a very 
simple way to regularize the matrix $\lambda$, as discussed in  the main text. 
In the case of odd $\bar{N}$ it will be shown later that there exists  always an  eigenvector  of $\lambda$  with vanishing  eigenvalue,  but the decomposition remains possible, as $\lambda_{MO}$ will contain at least one vanishing row and corresponding column.
  In the following we define that an eigenvector is singular if it corresponds 
  to a vanishing eigenvalue, as in the odd $\bar N$ case. 

It would be ideal for this calculation to use a very robust and stable diagonalization routine to maintain machine accuracy for the MOs. Unfortunately these routines are not commonly  available for  antisymmetric matrices and thus several mathematical transformations are necessary 
to map our task to a sequence of more  commonly used or at least easily 
available algorithms. 

A generic $\bar N \times \bar N$ antisymmetric matrix is written in  the following way: 
\begin{equation}
  \lambda=
  \left (
  \begin{matrix} 
    0 & a_{1,2} & a_{1,3} &  \cdots & a_{1,\bar N} \\ 
    -a_{1,2} & 0 & a_{2,3} &  \cdots & a_{2,\bar N} \\
    -a_{1,3} & -a_{2,3} & 0 &  \cdots & a_{3,\bar N} \\
    \vdots & \vdots & \vdots & \ddots & \vdots \\
    -a_{1,\bar N} & -a_{2,\bar N} & -a_{3,\bar N} &  \cdots & 0  
  \end{matrix}
  \right ).
  \label{startlambda}
\end{equation}
The first step is to transform $\lambda$ in a tridiagonal antisymmetric real matrix. This operation is implemented in the subroutine zsktrd (dsktrd) contained in the PFAPACK library \cite{Wimmer:2012:A9E:2331130.2331138}.  The use of the Householder algorithm allows us to decompose the generic matrix $\lambda$ as
\begin{equation}
  \lambda=U^*_{1}\lambda_{Tr} U^\dag_{1},
  \label{tridiag}
\end{equation}
where $U_1$ is the transformation matrix output of the algorithm, while $\lambda_{Tr}$ is a tridiagonal real antisymmetric matrix written in the standard tridiagonal form
\begin{equation}
  \lambda_{Tr}=
  \left (
  \begin{matrix} 
    0 & b_1  & 0 &  \cdots & 0 \\ 
    -b_{1} & 0 & b_2 &  \cdots & 0  \\
    0 & -b_2 & 0 &  \cdots & 0 \\
    \vdots & \vdots & \vdots & \ddots & \vdots 
  \end{matrix}
  \right ).
  \label{UTR}
\end{equation}

Thus we can multiply the matrix $\lambda_{Tr}$ for the imaginary unit $i$,   yielding a 
more conventional tridiagonal hermitian matrix $\lambda_{iH}$,  defined by purely imaginary matrix elements. 

We highlight that it is possible to map the matrix $\lambda_{iH}$ into a real hermitian matrix via a unitary transformation and use the appropriate  LAPACK routine for its fast diagonalization. This procedure is well known and will be discussed later.

At this point,  we can use the spectral theorem for Hermitian matrices  to decompose the matrix $\lambda_{iH}=\psi \lambda_{diag} \psi^\dag$, where $\lambda_{diag}$ is a diagonal matrix containing in its diagonal part the  real eigenvalues $a_i$  of 
$\lambda_{iH}$ and $\psi$ is the unitary matrix,  where each column is given  by   the eigenvector 
, in principle complex, corresponding  to each eigenvalue,  in the chosen order.
This decomposition implies: 
\begin{equation}
  \lambda=-i U_{1}^*  \psi \lambda_{diag} \psi^\dag  U_{1}^\dag.
  \label{decomposition}
\end{equation} 
\\
However,  since the  matrix $\psi$ is generally complex and $\psi^\dag \ne \psi^T$, 
some manipulation is necessary if we want to satisfy the skew-symmetry property of $\lambda$,  
in an easy and transparent  way.

If we consider one eigenvector $\bar{\psi}_j$ associated to an eigenvalue $a_j>0 $ we have that
\begin{equation}
  \lambda_{iH}\bar{\psi}_j= i \lambda_{Tr}\bar{\psi}_j= a_j \bar{\psi}_j,
  \label{cc}
\end{equation} 
the complex conjugate of this expression is 
\begin{equation}
  -i \lambda_{Tr}\bar{\psi}^*_j= -a_j \bar{\psi}_j^*,
\label{cc2}
\end{equation} 
where we have used that both  $\lambda_{Tr}$  and the eigenvalues $a_j$ are real.
This means that if $\bar{\psi}_j$ is an eigenvector of $\lambda_{iH}$ relative to the eigenvalue $a_j$, then $\bar{\psi}_j^*$ is an eigenvector corresponding to the eigenvalue $-a_j$ and thus 
orthogonal to $\bar{\psi}_j$ because of the  orthogonality between eigenvectors of an Hermitian matrix corresponding to 
different eigenvalues $\pm a_j$. We can thus easily verify,  by using  the relations given in  Eq.~(\ref{cc}) and Eq.(\ref{cc2}), the  following 
simple equations:
\begin{eqnarray}
  \lambda_{iH}(\bar{\psi}_j+\bar{\psi}_j^*)=  a_j (\bar{\psi}_j-\bar{\psi}_j^*) \\
   \lambda_{iH}(\bar{\psi}_j-\bar{\psi}_j^*)=  a_j (\bar{\psi}_j+\bar{\psi}_j^*)
  \label{cc-pairs}
\end{eqnarray}\footnote{The same argument holds if the eigenvalue $a_j$ corresponds to 
$p>1$  degenerate eigenvectors. The mentioned orthogonality property of Hermitian matrix  
eigenvectors leads to the straightforward 
definition of  $p$ pairs of mutually orthonormal real ones used for the present decomposition 
with a   block diagonal matrix, where each $2\times 2$ block corresponds to one of the  
$p$ degenerate eigenvectors.}.
In this way  we can define pairs of {\em real} orthogonal vectors $\bar{\psi}_j^1=\sqrt{2} \  \Re ({\bar{\psi}_j})$ and $\bar{\psi}_j^2=\sqrt{2} \ \Im (\bar{\psi}_j)$  such that
\begin{eqnarray}
  \lambda_{iH} \bar{\psi}_j^1=i a_j \bar{\psi}_j^2\\
  \lambda_{iH} \bar{\psi}_j^2=-i a_j \bar{\psi}_j^1.
\end{eqnarray}
Once we have identified all the pairs corresponding to all positive  eigenvalues $a_j>0$ 
we can write the unitary matrix $\bar{\psi}$ 
that is now real, by  adding  the remaining eigenvectors (that can be also chosen real as shown in Subsec.\ref{sec:singular}) with vanishing eigenvalues in the remaining rightmost columns. In this way we can finally define a unitary real matrix $\bar \psi$ yielding $\lambda_{Tr} =-i \lambda_{iH} = \bar \psi
\lambda_{MO} \bar \psi^T$ where $\lambda_{MO}$  is defined in Eq.~(\ref{diaglambda}) and therefore 
by using Eq.(\ref{tridiag})
\begin{equation}
  \lambda= U_{1}^* \bar  \psi \lambda_{MO} \bar \psi^T  U^\dag_{1}.
  \label{matrix}
\end{equation}
which represents the desired decomposition because  the product of two unitary matrices 
$U^*= U_1^* \bar \psi  $ remains a unitary matrix and its transponce $U^\dag$ coincides with  $\bar \psi^T 
U_1^\dag$, yielding $\lambda= U^* \lambda_{MO} U^\dag$,  that is the purpose of this appendix.

\subsection{Triangular hermitian matrices: a mapping from imaginary to real}

In order to use the LAPACK routines for the diagonalization we have to map the tridiagonal fully imaginary hermitian matrix $\lambda_{iH}$, defined only (the  diagonal elements are zero to fullfill hermitianity) by  its upper diagonal elements  $ i b_j$ with  $b_j$ real for $j=1,2,\cdots \bar N-1$,
into a tridiagonal real symmetric matrix $\lambda_{R}$. We can implement this mapping by applying a unitary transformation  to the matrix $\lambda_{iH}$. For this purpose we introduce the following transformation described by the matrix $U_2$
\begin{equation}
\lambda_R=U_2^\dagger  \lambda_{iH} U_2.
  \label{unitary}
\end{equation}
The matrix $U_2$ is a complex diagonal matrix defined as
\begin{equation}
  U_2=
  \left (
  \begin{matrix}
     e^{i\phi_1} & 0 & \cdots & 0 \\
     0 & e^{i\phi_2} & \cdots & 0 \\
     \vdots & \vdots & \ddots & \vdots \\
     0 & 0 & \cdots & e^{i\phi_{\bar N-1}}
  \end{matrix}
  \right ).
  \label{U3}
\end{equation}
 The explicit calculation of the right-hand  side of the 
eq.~(\ref{unitary}) gives
\begin{equation}
  \lambda_{R}=
  \left (
  \begin{matrix} 
    0 & i b_{1} e^{i(\phi_2-\phi_1)} & 0 &  \cdots & 0 \\ 
    -i b_{1}e^{i(\phi_1-\phi_2)} & 0 &i  b_2e^{i(\phi_3-\phi_2)} &  \cdots & 0  \\
    0 & -i b_2 e^{i(\phi_2-\phi_3)}& 0 &  \cdots & 0 \\
    \vdots & \vdots & \vdots & \ddots & \vdots 
  \end{matrix}
  \right ).
  \label{op3}
\end{equation}
By setting the  imaginary  units $\pm i=\exp( \pm i {\pi \over 2})$ 
(when not  exponentiated in the previous equation), we can easily impose that all the phase factors cancel in all the corresponding 
matrix elements of $\lambda_R$
with the  choice:  
\begin{equation}
  \phi_j = - \frac \pi 2 (j-1),
  \label{phi}
\end{equation}
that therefore implies that $\lambda_R$, with the above definition, 
is a real symmetric matrix.
At this point we can  diagonalize the matrix $\lambda_R$ by means of a real unitary matrix $U_R$,
that is the output of a standard  LAPACK diagonalization routine of tridiagonal real matrices (e.g. 
dstevx for double precision arithmetic).
In this way $\lambda_{iH}$ can be diagonalized as  $\lambda_{iH}=U_2 U_R \lambda_{diag} U_R^T U_2^\dag$ 
where $\lambda_{diag}$ is a diagonal matrix contaning  the corresponding 
eigenvalues of the LAPACK diagonalization.

\subsection{Singular  eigenvectors}
\label{sec:singular}

Within this formulation it  is also particularly easy  to compute 
all the real singular eigenvectors
of $\lambda_{iH}$ corresponding to the possible 
vanishing eigenvalues. They were used in this appendix  to complete the columns 
of the unitary real matrix $\bar \psi$.
From the outcome of the   previous subsection any eigenvector $\phi^j_k$ 
of $\lambda_{iH}$ 
can be obtained by applying the diagonal matrix $U_2$ to a real eigenvector 
$\bar \phi^j_k$ of $\lambda_R$, 
namely  $\phi^j_k=\bar \phi^j_k \exp( - i {\pi \over 2} (k-1))$, implying that 
even $k-$components are purely imaginary and odd $k-$components are purely real.
Then it is immediate to realize that if $\bar \phi^j_k$ corresponds to 
a singular eigenvector of $\lambda_R$ 
also $\Re(\phi^j_k)$ and $\Im(\phi^j_k)$ ( and therefore 
also  $ \Re(\phi^j_k) + \Im(\phi^j_k)$)  correspond to singular 
eigenvectors or at most null vectors (not  both) of $\lambda_{iH}$ 
because this matrix is purely imaginary and the complex conjugation 
of a singular eigenvector is  again a singular eigenvector by Eq.~(\ref{cc}) 
and Eq.(\ref{cc2}) with $a_j=0$.

Then it follows that all the orthogonal eigenvectors $\bar \phi^j$ (output of dstevx)  corresponding 
to the zero eigenvalues of the matrix $\lambda_R$ can  be used to define 
the corresponding real singular eigenvectors corresponding to the matrix $\lambda_{iH}$, i.e.:
\begin{equation}
\tilde \phi^j_k = \Re(\phi^j_k) + \Im(\phi^j_k)= \left\{
\begin{array}{cc}
 (-1)^{(k-1)/2}  \bar\phi^j_k & k ~ {\rm odd} \\
 (-1)^{k/2}  \bar\phi^j_k & k ~ {\rm even} \\
\end{array} 
\right.
\end{equation}
that are explicitly real, and orthogonal each other because $\sum_k \tilde \phi^j_k 
\tilde \phi^l_k=\sum_k \bar \phi^j_k \bar \phi^l_k= \delta_{l,j}$.
They are also orthogonal to all the  other pairs of non singular  
eigenvectors  
 because of the orthogonality property of eigenvectors of an Hermitian matrix $\lambda_{iH}$, that we have already used in the previous section.

\end{document}